\RequirePackage{fix-cm}
\documentclass[smallextended]{svjour3}       
\smartqed  
\usepackage{wrapfig}
\usepackage{graphicx}%
\usepackage{multirow}%
\usepackage{array}
\usepackage{fontawesome}
\usepackage{tikz}
\usepackage{tabularx}
 \usepackage{natbib}
\usepackage{appendix}%
\usepackage{xcolor}%
\usepackage{textcomp}%
\usepackage{manyfoot}%
\usepackage{booktabs}%
\usepackage{amssymb}
\usepackage{listings}%
\usepackage{fancyhdr}
\usepackage{fancybox}
\usepackage[utf8]{inputenc}

\usepackage{enumerate}
\usepackage{enumitem}
\usepackage{url}
\usepackage{rotating}
\fancypagestyle{plain}{
    \fancyhf{}
    \fancyhead[LE,RO]{\thepage}
}
\begin{document}

\title{Navigating Fairness: Practitioners' Understanding, Challenges, and Strategies in AI/ML Development
}
\titlerunning{Navigating Fairness...}

\author{Aastha Pant*        \and  Rashina Hoda  \and Chakkrit Tantithamthavorn \and  Burak Turhan 
}

\authorrunning{Pant et al.}
          
\institute{A. Pant \at
              Department of Software Systems and Cybersecurity, Monash University, Melbourne, Australia \\
              \email{aastha.pant@monash.edu}           
           \and
           R. Hoda \at
              Department of Software Systems and Cybersecurity, Monash University, Melbourne, Australia \\
              \email{rashina.hoda@monash.edu}
              \and
                  C. Tantithamthavorn \at
              Department of Software Systems and Cybersecurity, Monash University, Melbourne, Australia \\
              \email{chakkrit@monash.edu}
              \and
                 B. Turhan \at
              Faculty of Information Technology and Electrical Engineering, University of Oulu, Oulu, Finland \\
              \email{burak.turhan@oulu.fi}
}

\date{Received: date / Accepted: date}

\maketitle

\begin{abstract}
The rise in the use of AI/ML applications across industries has sparked more discussions about the fairness of AI/ML in recent times. While prior research on the fairness of AI/ML exists, there is a lack of empirical studies focused on understanding the perspectives and experiences of AI practitioners in developing a fair AI/ML system. Understanding AI practitioners' perspectives and experiences on the fairness of AI/ML systems is important because they are directly involved in its development and deployment and their insights can offer valuable real-world perspectives on the challenges associated with ensuring fairness in AI/ML systems. We conducted semi-structured interviews with 22 AI practitioners to investigate their \emph{understanding} of what a `fair AI/ML' is, the \emph{challenges} they face in developing a fair AI/ML system, the \emph{consequences} of developing an unfair AI/ML system, and the \emph{strategies} they employ to ensure AI/ML system fairness. We developed a framework showcasing the relationship between AI practitioners' \emph{understanding} of `fair AI/ML' system and (i) their \emph{challenges} in its development, (ii) the \emph{consequences} of developing an unfair AI/ML system, and (iii) \emph{strategies} used to ensure AI/ML system fairness. By exploring AI practitioners' perspectives and experiences, this study provides actionable insights to enhance AI/ML fairness, which may promote fairer systems, reduce bias, and foster public trust in AI technologies. Additionally, we also identify areas for further investigation and offer recommendations to aid AI practitioners and AI companies in navigating fairness.
\keywords{artificial intelligence \and machine learning \and AI fairness \and AI practitioners \and interviews}
\end{abstract}

\section{Introduction}\label{Intro}

In recent years, the use of AI/ML systems has become widespread across various domains, including recruitment, legal proceedings, credit risk forecasting, and admission processes \citep{mehrabi2021survey}. `Fairness' has been a subject of study in Software Engineering (SE) research for some time, predating the recent surge in AI/ML applications \citep{finkelstein2008fairness}. At the same time, the importance of `fairness' of AI/ML systems has been highlighted by several real-world incidents in recent years \citep{majumder2023fair}. For example, there have been fairness issues in AI/ML systems such as Google's ML algorithm exhibiting gender bias against women by more frequently associating men with Science, Technology, Engineering, and Mathematics (STEM) careers \citep{prates2020assessing}; Amazon’s AI-powered recruitment tool that was gender-biased as it preferred male candidates over female candidates based on their resumes \citep{Amazon}; a risk score predicting algorithm exhibiting significant bias against African Americans, revealing a higher error rate in predicting future criminals \citep{Machine}; gender bias in \emph{Google} \citep{caliskan2017semantics} and \emph{Bing} translators \citep{johnson2022fairkit}. Widespread cases of software displaying unfair behavior, particularly regarding protected attributes such as gender \citep{caliskan2017semantics} and race \citep{Machine}, underscore the necessity of prioritising `fairness' in the development of AI/ML systems, as these instances lead to unacceptable consequences disproportionately affecting users in minority or historically disadvantaged groups.

The widespread adoption of AI/ML systems across different domains has raised concerns about fairness, leading to increased research and the development of guidelines and policies. Major tech companies like Google \citep{Google}, Microsoft \citep{Microsoft}, IBM \citep{IBM}, and various countries/ continents, including Australia \citep{Australisethics} and Europe \citep{EUethics}, have defined `fairness' as a guiding principle for AI practitioners in developing a fair AI/ML system. The essence of the `fairness' principle for these countries/continents and tech companies is centered around developing an inclusive AI/ML system that does not discriminate against any specific individuals, groups, or communities. Along with that, several software and tools have also been developed such as IBM's AI Fairness 360 \citep{AI360}, LinkedIn's Fairness Toolkit (LiFT) \citep{vasudevan2020lift}, and fairness checklists like Deon \citep{Deon}, Microsoft's AI Fairness Checklist \citep{Microsoft_checklist}, IBM's AI FactSheets \citep{AIFact} and many more to aid AI practitioners in developing a fair AI/ML system. The extensive research in the field of AI/ML system fairness covers various aspects, including the proposal of methods and frameworks \citep{johnson2022fairkit, zhang2023fairness}, aimed at aiding AI practitioners in the design and development of a fair AI/ML system or mitigating fairness-related issues in them. Despite the development of numerous tools, frameworks, guidelines, and policies for AI/ML fairness, issues persist. Our recent survey study also showed that most AI practitioners discussed facing challenges in developing \textbf{fair} AI/ML systems because of their own biased nature \citep{pant2023ethics}. 

The predominant focus has been on introducing guidelines, and policies, and developing tools for AI practitioners to enhance the development of a fair AI/ML system. Given that human society is diverse in terms of cultures, experiences, and viewpoints, AI teams must reflect this diversity to effectively create fair and impactful technologies \citep{xavier2024biases}. Therefore, understanding the perspectives and experiences of these practitioners who are actively involved in AI/ML system development is equally crucial. This deeper understanding can play a pivotal role in uncovering real-world challenges encountered during the development process. Such awareness can help to devise solutions that can directly address practical needs and concerns identified by practitioners, thereby aiding in the development of fair AI/ML systems and mitigating societal inequalities \citep{holstein2019improving}. Understanding what `fairness' means in the context of AI/ML from practitioners' perspectives may help policymakers create better regulations that tackle real-world issues and promote ethical AI deployment. This approach may enhance inclusivity, reduce discrimination risks, and boost public trust in AI systems \citep{dankloff2024analysing}. Ultimately, it creates a digital environment where AI enhances societal well-being. A recent study has also reported that most studies on AI/ML system fairness are conceptual and focused on technical aspects, highlighting the importance and need for research on the social/human aspects of AI \citep{xivuri2021systematic}. 

Therefore, considering the importance of understanding the overall perspectives and experiences of AI practitioners in the development of a fair AI/ML system, as emphasised in the literature, and taking into account the identified research gap \citep{xivuri2021systematic}, we were interested in addressing this gap by conducting an empirical study with AI practitioners\footnote{The term `AI practitioners’ in our study includes AI/ML developers, AI engineers, AI/ML experts, and AI/ML/ data scientists involved in the design and development activities of AI/ML systems. The terms ‘AI practitioners’ and `practitioners’ are used interchangeably throughout our study.}. We conducted \textbf{semi-structured interviews} with 22 AI practitioners to explore four aspects: (i) AI practitioners' understanding of `fair AI/ML', (ii) their challenges in fair AI/ML development, (iii) consequences of developing an unfair AI/ML system, and (iv) their strategies\footnote{The term `strategy' in our study refers to practical, day-to-day approaches aimed at ensuring the fairness of AI/ML, rather than encompassing a broader, overarching plan or approach intended for achieving long-term goals.} to ensure the fairness of an AI/ML system. The study aims to answer the following four research questions (RQs):

\textbf{RQ1. What do AI practitioners understand by `fair AI/ML'?}\\
To address RQ1, we explicitly asked AI practitioners about their understanding of `fair AI/ML'. This approach was chosen to investigate how `fairness' is understood by AI practitioners in the context of AI/ML.

\textbf{RQ2. What challenges do AI practitioners face in developing a fair AI/ML system and what are the factors that lead to those challenges?}\\
To address RQ2, we inquired with AI practitioners about the overall challenges they encounter in developing a fair AI/ML system, drawing insights from their experiences. Additionally, we explored the underlying factors contributing to those challenges. 

\textbf{RQ3. What do AI practitioners perceive as the consequences of developing an unfair AI/ML system?}\\
To address RQ3, we asked AI practitioners to share their perceptions of the consequences associated with developing an unfair AI/ML system. The question went beyond inquiring about their experiences, and also seeking their overall perspective on the consequences of developing an unfair AI/ML system. 

\textbf{RQ4. What strategies do AI practitioners use in ensuring the fairness of an AI/ML system?}\\
To address RQ4, we asked AI practitioners about their practical, day-to-day approaches derived from their experience in ensuring the fairness of the AI/ML system they develop. 

We used \emph{Socio-Technical Grounded Theory (STGT) for data analysis} \citep{hoda2021socio} to analyse the qualitative data. The main contributions of this study are:
\begin{itemize}
    \item We investigated what AI practitioners \emph{understand} by ‘fair AI/ML’.
    \item We identified the \emph{challenges} faced by AI practitioners in developing a fair AI/ML system and the factors leading to those challenges.
    \item We identified the \emph{consequences} of developing an unfair AI/ML system perceived by AI practitioners.
    \item We explored the \emph{strategies} used by AI practitioners to ensure the fairness of the AI/ML system they developed.
    \item We developed a \emph{framework} illustrating the relationship between AI practitioners’ \emph{understanding} of ‘fair AI/ML’ and (i) \emph{challenges} faced in developing a fair AI/ML system, (ii) the \emph{consequences} of developing an unfair AI/ML system, and (iii) \emph{strategies} for ensuring the fairness of an AI/ML system. 
\item We formulated a set of recommendations for AI practitioners and AI companies to assist them in the development of fair AI/ML systems based on the empirical findings. 
\end{itemize}

\section{Background and Motivation} \label{sec:Related work} 
\subsection{Definition and Approaches on `AI/ML fairness'}
In recent years, the concept of `fairness' in AI has gained significant attention. Leading software companies such as Microsoft, Google, and IBM have either outlined principles or recommended practices to guide practitioners in developing fair AI systems. For instance, Microsoft has defined `fairness' as \emph{``AI systems should treat all people fairly"} \citep{Microsoft}. Likewise, IBM emphasises the importance of minimising bias and promoting inclusive representation in AI development \citep{IBM}. Meanwhile, Google recommends concrete steps for fair AI, including setting clear goals for fairness, using representative datasets, checking systems for unfair biases, and analysing system performance \citep{Google}. In addition to companies, various countries and continents have their definitions of the term `fairness' in the context of AI. For example, Australia’s AI Ethics Principles defined the `fairness' principle as \emph{``AI systems should be inclusive and accessible, and should not involve or result in unfair discrimination against individuals, communities or groups"} \citep{Australisethics}. Similarly, the European Commission defined `Diversity, non-discrimination and fairness’ in AI as, \emph{“Unfair bias must be avoided, as it could have multiple negative implications, from the marginalisation of vulnerable groups to the exacerbation of prejudice and discrimination. Fostering diversity, AI systems should be accessible to all, regardless of any disability, and involve relevant stakeholders throughout their entire life circle"} \citep{EUethics}.

Along with that, over the last 13 years, there has been extensive research on AI/ML fairness \citep{friedler2019comparative}, and different tools, techniques, and methods to measure and mitigate fairness issues in AI/ML systems have been developed and evaluated. Major tech companies like Microsoft, Google, and IBM have developed software tools and techniques to enhance the development of fair AI/ML systems such as AI Fairness 360 \citep{AI360}, LinkedIn's Fairness Toolkit (LiFT) \citep{vasudevan2020lift}, and fairness checklists like Deon \citep{Deon}, Microsoft's AI Fairness Checklist \citep{Microsoft_checklist}, IBM's AI FactSheets \citep{AIFact}. Furthermore, researchers have developed a variety of methods and frameworks intending to enhance the development of fair AI/ML systems, including fairness checklists \citep{madaio2020co}, frameworks \citep{vasudevan2020lift, d2020fairness}, and fairness evaluation and comparison toolkit \citep{johnson2022fairkit}. 

\subsection{Review studies on AI/ML fairness}
In addition to defining the concept of \emph{AI fairness}, several review studies have been conducted in the area of \emph{fairness} of the AI/ML systems. For example, studies have been conducted to explore and review the definition of \emph{fairness} focused on various aspects such as ML algorithmic classification \citep{verma2018fairness}, the widely used definition in ML \citep{mehrabi2021survey, chouldechova2018frontiers} and political philosophy \citep{binns2018fairness}. Likewise, studies have also been conducted to compare the historical and current perspectives of fairness in ML \citep{hutchinson201950}. Studies have also focused on reviewing the challenges and methodologies related to AI fairness. For example, \citet{chen2023ai} conducted a literature review of 59 articles to explore the challenges in ensuring AI fairness and the strategies to improve fairness in AI systems. Likewise, \citet{xivuri2021systematic} performed a systematic literature review (SLR) with 47 articles, examining AI algorithm fairness across research methods, practices, sectors, and locations. Their findings revealed a predominance of conceptual research, primarily emphasising the technical aspects of narrow AI, and highlighted a notable gap in research, specifically the lack of research on the social and human aspects of AI. \citet{pessach2022review} conducted a review study on ML fairness focusing on exploring the causes of algorithmic bias, common definition, and measures of fairness. \citet{caton2020fairness} conducted a review study to provide an overview of different approaches used to increase the fairness of ML systems. \citet{pagano2023bias} conducted a systematic review to explore various aspects like datasets, fairness metrics, tools, and identification and mitigation methods of mitigating bias and unfairness in ML systems. Likewise, \citet{bacelar2021monitoring} provided an overview of various measurement methods of bias and fairness in ML models, in their review study. \citet{wan2023processing} provided a review of the currently available mitigation techniques of in-procession fairness issues in ML models. Review studies have also been conducted to address the fairness issues, the causes of biases in AI, and their consequences in the medical domain \citep{ueda2024fairness}. Likewise, \citet{wang2023multidisciplinary} conducted a review of 95 articles to explore similarities, and differences in the understanding of fairness, influencing factors, and potential solutions for fairness integration in medical AI.    

The emphasis of major tech companies and nations has largely been on working to define the concept of `fairness' and develop diverse tools and techniques to assist AI practitioners in enhancing the development of fair AI/ML systems. Despite these efforts, our recent survey study showed that most participants reported the challenges in developing \textbf{fair} AI/ML systems due to their own biased nature \citep{pant2023ethics}. Given that most studies on AI/ML fairness are conceptual and focused on technical aspects, and considering the highlighted importance and need for research on the social/human aspects of AI in the literature \citep{xivuri2021systematic}, we were interested in exploring the perceptions and experiences of AI practitioners regarding fair AI/ML systems. Investigating AI practitioners' perceptions and experiences in developing a fair AI/ML system can assist in understanding the real-world challenges associated with fair AI/ML system development. Furthermore, it can aid in devising solutions to address their practical needs and concerns in developing a fair AI/ML system. 

\section{Research Methodology} \label{sec:Methodology}
\begin{figure}[htpb]
    \centering
    \includegraphics[width=\textwidth] {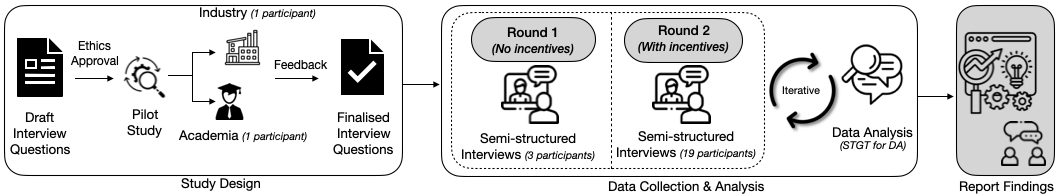}
    \caption{Overview of the research methodology of our study}
    \label{fig:methodology}
\end{figure}

Our study aimed to investigate the perspectives and experiences of AI practitioners in developing a fair AI/ML system. Figure \ref{fig:methodology} shows the overview of the research methodology of our study. 

\subsection{Study Design}
We conducted a semi-structured interview-based study which commonly allows researchers to study the complexities of human behavior such as motivation, communication, and understanding to obtain rich and informative results \citep{seaman1999qualitative}. We conducted semi-structured interviews focusing on AI/ML fairness and the findings are divided into two primary categories. A smaller part of the findings which revolves around AI practitioners' perspectives and experiences on `AI/ML bias', has been accepted for publication in IEEE Software \citep{pant2024aimlpractitionersthinkaiml}. The larger part (core part) of the findings related to `fair AI/ML', is presented in this paper. The complete interview protocol is provided in Appendix \ref{Appendix A}.

Basically, we gathered AI practitioners' insights on their \emph{understanding} of `fair AI/ML', the \emph{challenges} they face in developing a fair AI/ML system, \emph{consequences} of developing an unfair AI/ML system perceived by them, and the \emph{strategies} they take to ensure fairness of an AI/ML system. Interview planning spanned from July 2023 to October 2023. Throughout this period, tasks included defining interview objectives, refining the interview protocol through iterative processes, and prioritising crucial interview questions. Consequently, a semi-structured interview protocol with two sections was developed.  

\subsubsection{Participant Information} The first section of the interview protocol was formulated to gather participants' demographic information, including their name, email, gender, age, country of residence, and educational qualifications. Employment details such as job titles, and involvement in AI/ML system development activities were also collected. We used a pre-interview questionnaire to gather the participants' demographic information. Participants were also asked to provide details of their work experience in the area of AI/ML system development, and those without experience were not included in the study. Each participant included in our study has at least some experience in the area of AI/ML system development. Using the \textit{Qualtrics} platform, we created the pre-interview questionnaire and advertised it as an anonymous survey link following the receipt of necessary ethics approval (Reference Number: 38991). The pre-interview questionnaire can be found in Appendix \ref{Appendix A} - Section A. 

\subsubsection{Understanding Participants' Perspectives and Experiences in Developing a Fair AI/ML System}
The second section of the interview protocol was designed to gather insights into participants' perspectives and experiences in the development of a fair AI/ML system. At the start of the interview, we asked participants if they were familiar with the term `fair AI/ML', and if they had encountered any fairness-related cases while developing AI/ML systems. Only those who had experience with fairness-related cases in their professional work were recruited for the interview.

Our focus was specifically on investigating AI practitioners' \textit{understanding} of `fair AI/ML'. We did not provide a predefined definition of `fairness' to participants and explicitly inquired about their understanding, aiming to assess their perspectives independently. The two key reasons for this design choice include: (i) as mentioned in section \ref{sec:Related work}, there is no universal definition of `fairness' in AI—different countries and tech companies have their own definitions of `fairness' and (ii) this approach aimed to evaluate participants' natural interpretations, avoiding influence from a predetermined definition. We also aimed at identifying their \textit{challenges} in developing fair AI/ML and the factors leading to those challenges, understanding the \textit{consequences} of developing an unfair AI/ML system from the participants' perspective, and exploring the \textit{strategies} they employ to ensure fairness of an AI/ML system. To ensure we captured real-world experiences, we asked participants for real-world examples and experiences during the interview. The interview questions can be found in Appendix \ref{Appendix A} - Section B.

\subsubsection{Pilot study}
After designing the interview protocol, we executed a pilot study, engaging two AI practitioners—one from industry and another from academia—identified through our professional networks. The purpose was to confirm the clarity and understandability of the interview questions, assess the time required to complete the study and gather feedback for enhancing the interview process. Both participants possessed expertise in AI/ML system development. Taking into account their feedback, we made slight modifications to the interview questions to enhance clarity, ultimately finalising the interview protocol.

\subsection{Interview Sampling and Data Collection}
We used purposive sampling in our study to select the participants \citep{baltes2022sampling}. By using this method, we were able to specifically target our desired group of participants, namely AI practitioners involved in AI/ML system development activities. 

We conducted data collection in two rounds. In the first round, participation was voluntary. After we got the ethics approval, we advertised our study on social media platforms such as LinkedIn and Twitter, as well as within our professional networks. We specifically targeted AI practitioners engaged in AI/ML system development activities. In the first round, we received interest from only 3 candidates for participating in our study. So, after obtaining ethics approval, we decided to conduct a second round of data collection and introduced a reward— an AUD 50 gift card voucher— to incentivise participation. The second round, advertised again on social media like LinkedIn and Twitter with mention of the reward, resulted in responses from 19 suitable candidates, bringing the total number of participants to 22. Since our goal was to recruit participants with some experience in AI/ML system development, we incorporated two employment-related questions, inquiring about their years of experience in the field and their level of involvement in various job responsibilities. Participants received a reward of AUD 50 upon the completion of data collection. Since we advertised our study on social media, we obtained responses from various countries worldwide, as illustrated in Table \ref{table:Demo}. Since we did not favour specific countries, the responses were spread out across different regions. We obtained a majority of the responses from Australia (13), followed by the responses from other countries like Nepal (3), Israel (1), Japan (1), USA (1) etc. We present an in-depth analysis of the participants’ demographics in Section \ref{Sec5:Demo}. 

We gathered qualitative data through semi-structured interviews with 22 AI practitioners experienced in AI/ML system development. All interviews were conducted online using Zoom and were audio-recorded. Each interview lasted between 40 and 45 minutes.

\subsection{Data Analysis}
\begin{figure}[htpb]
    \centering
    \includegraphics[width=\textwidth] {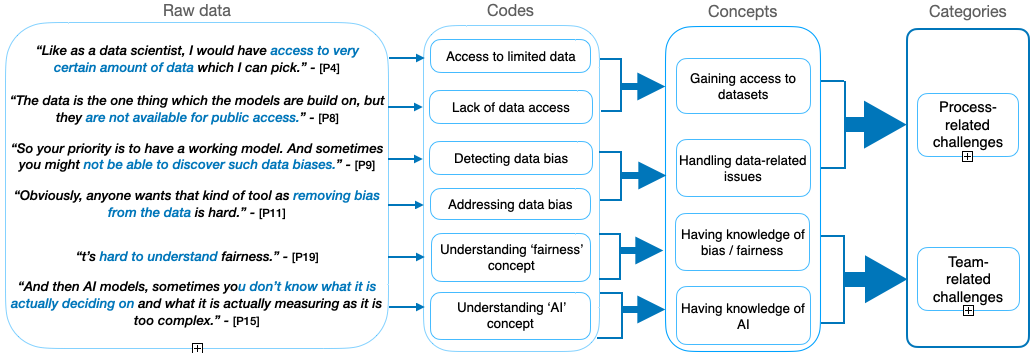}
    \caption{Examples of STGT analysis \citep{hoda2021socio} applied to qualitative data on the \emph{challenges} in developing a fair AI/ML system.}
    \label{fig:stgt}
\end{figure}

In our study, qualitative data was gathered via semi-structured interviews, and consequently, a qualitative approach was employed for data analysis. \emph{Socio-Technical Grounded Theory (STGT) for data analysis} was used to analyse the data, as it is particularly suitable for analysing open-ended data and gaining insights within socio-technical contexts \citep{hoda2021socio}. After obtaining consent from each participant, we transcribed the data. The data collection and analysis phases involved an iterative process as shown in Figure \ref{fig:methodology}. Initially, we analysed the data from 13 participants using \emph{open coding} approach to develop concepts and categories, involving constant comparison of diverse open-text responses \citep{hoda2021socio}. We performed inductive open coding within the RQs. For example, to answer our RQ2  which is, \textit{What challenges do AI practitioners face in developing a fair AI/ML system and what are the factors that lead to those challenges?}, initially, we gathered qualitative data from 13 participants by asking them, {\emph{``Based on your professional experience, do you face any challenges in developing a fair AI/ML system? (If yes), what challenges do you face? What do you think are the factors leading to those challenges?"}} We developed codes using the open-coding approach in open-text answers as shown in Figure \ref{fig:stgt}. For instance, codes like \emph{`access to limited data'} and \emph{`lack of data access'} were identified through open coding. Subsequently, we engaged in constant comparison of these codes to continually compare them, leading to the recognition of patterns among them. For instance, upon reviewing the codes mentioned above, we identified a common pattern related to the challenge of accessing datasets required in the development of a fair AI/ML system. We combined these two codes to develop a concept of \emph{`gaining access to datasets'}. 

Using the same constant comparison approach for other codes, we derived concepts such as \emph{`balancing ideal vs real'}, \emph{`handling data-related issues'}, and \emph{`following policies and regulations'}. We again constantly compared these concepts with one another and developed distinct categories. In this context, these four concepts shared a challenge associated with the \emph{process} of developing a fair AI/ML system, leading us to establish a category known as \emph{`process-related challenges'}.

Likewise, we identified multiple codes and concepts addressing the challenges associated with the resources required for developing a fair AI/ML system. This process led to the development of another high-level category, namely, \emph{`resource-related challenges'}. In this way, we established a total of three categories encapsulating the challenges faced by AI practitioners in developing a fair AI/ML system, namely, \textbf{process-related challenges}, \textbf{resource-related challenges}, and \textbf{team-related challenges}. Detailed information on these challenges is provided in Section \ref{Sec5:Findings}.

Building on the primary findings from the initial analysis, we collected data from the remaining 9 participants, focusing on key insights from the first round. This data was analysed using \emph{targeted coding}, which involves generating codes that align with the concepts and categories identified in the initial stage, following STGT guidelines \citep{hoda2021socio}.

All four authors were involved in designing the interview questionnaire. However, the first author led the data analysis with detailed feedback from the second author and regular feedback from the third and fourth authors. After the qualitative data were analysed, the results, including codes, concepts, and categories, were shared and discussed among all authors, who collectively contributed to presenting the findings.

\begin{figure}[htpb]
    \centering
    \includegraphics[width=\textwidth] {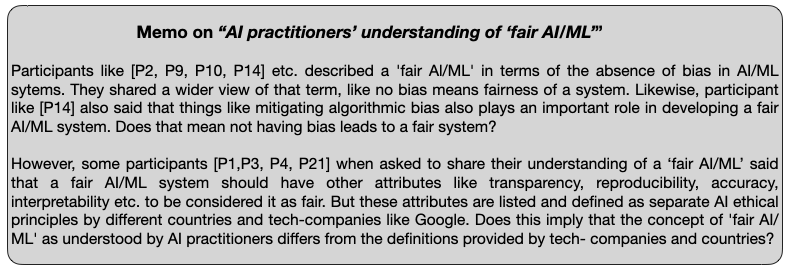}
    \caption{An example of a memo on AI practitioners' understanding of `fair AI/ML'}
    \label{fig:memo}
\end{figure}

The \emph{STGT for data analysis} encompasses steps of open coding, targeted coding, constant comparison, and memoing. \emph{``Basic memoing is the process of documenting the researcher’s thoughts, ideas, and reflections on emerging concepts and (sub)categories and evidence-based conjectures on possible links between them"} \citep{hoda2021socio}. Consequently, we wrote memos to record significant insights and reflections discovered during the open coding and targeted coding activities. An illustration of a memo created for AI practitioners' description of `fair AI/ML', specifically `in terms of the absence of \emph{bias}' and `in terms of the presence of desirable \emph{attributes}', is provided in Figure \ref{fig:memo}. The discussion on the key insights derived from memoing is presented in Section \ref{sec:Insights}.

\section{Findings} \label{Sec5:Findings}
\subsection{Participants' Demographics} \label{Sec5:Demo}
We present the demographic information of the participants in this section. Table \ref{table:Demo} presents an overview of the participants' demographics based on their age, gender, country, education, work experience in AI/ML system development activities, and job title. We used identifiers such as P1, P2, P3, and so forth to represent the participants in our study.

\begin{table*}[htbp]
\centering
\caption{Demographics of the Interview Participants} \label{table:Demo}     
\footnotesize
\begin{tabular} {>{\raggedright\arraybackslash}p{0.5cm}>{\raggedright\arraybackslash}p{1cm}>{\raggedright\arraybackslash}p{1cm}>{\raggedright\arraybackslash}p{1.1cm}>{\raggedright\arraybackslash}p{2.3cm}>{\raggedright\arraybackslash}p{0.8cm}>{\raggedright\arraybackslash}p{2.3cm}}

 \hline\noalign{\smallskip}
P\_Id & Age Range (years) & Gender & Country & Education & AI/ML Exp. (years) & Job Title  
\\

\hline\noalign{\smallskip}

P1 & 20-25 & Man & Nepal & Bachelor & 1-2 & AI Engineer 
\\

P2 & 20-25 & Man & Australia & Bachelor & 0-1 & AI Engineer 
\\

P3 & 31-35 & Woman & Thailand & Ph.D. or higher & 2-5 & AI Research Scientist 
\\

P4 & 26-30 & Man & India & Bachelor & 5+ & Data Scientist 
\\

P5 & 26-30 & Man & Australia & Master & 5+ & AI Engineer 
\\

P6 & 26-30 & Man & Australia & Master & 1-2 & Data Scientist 
\\
P7 & 31-35 & Woman & Australia & Master & 1-2 & AI Engineer 
\\

P8 & 26-30 & Man & Nepal & Master & 1-2 & ML Engineer 
\\

P9 & 31-35 & Man & Australia & Ph.D. or higher & 5+ & Data Scientist 
\\

P10 & 46-50 & Man & Australia & Ph.D. or higher & 2-5 & ML Engineer 
\\

P11 & 26-30 & Man & Australia & Master & 1-2 & ML Engineer 
\\

P12 & 20-25 & Woman & Australia & Master & 1-2 & ML Engineer 
\\

P13 & 26-30 & Man & Australia & Master & 2-5 & ML Engineer 
\\

P14 & 31-35 & Man & Australia & Ph.D. or higher & 2-5 & Data Scientist 
\\

P15 & 46-50 & Man & Japan & Master & 2-5 & AI Engineer 
\\

P16 & 31-35 & Man & Australia & Master & 2-5 & ML Engineer 
\\

P17 & 26-30 & Man & Australia & Bachelor & 1-2 & ML Engineer 
\\

P18 & 31-35 & Woman & Australia & Master & 0-1 & ML Engineer 
\\

P19 & 31-35 & Man & Vietnam & Master & 2-5 & AI Engineer 
\\

P20 & 31-35 & Man & Israel & Master & 5+ & ML Expert 
\\

P21 & 20-25 & Man & Nepal & Bachelor & 2-5 & ML Engineer 
\\

P22 & 26-30 & Woman & USA & Master & 2-5 & Data Scientist 
\\

\noalign{\smallskip}\hline
\end{tabular}
\end{table*}

A total of 22 AI practitioners took part in our study including 17 men and 5 women. Moreover, the majority of participants (8 out of 22) fell into the age group of 26-30 years and 31-35 years each, while only 2 belonged to the age group of 46 to 50 years. In terms of experience, the majority (13 participants) had more than 2 years of experience whereas 9 participants had up to 2 years of experience in AI/ML system development. The geographical distribution indicated that the majority were from Australia (13 participants), with 3 participants from Nepal and 1 each from India, Japan, USA, Israel, Thailand, and Vietnam. Similarly, we inquired about the participants' job titles or roles within their companies. The majority of participants held the title of `ML Engineer' (9 out of 22), followed by `AI Engineer' (6 out of 22), `Data Scientist' (5 out of 22), and one participant each for `AI Research Scientist' and `ML Expert'. As our target interview participants were practitioners involved in AI/ML system development activities, we wanted to know the major AI/ML system development-related activities they were involved in. Among the 22 participants, the majority engaged in `Data cleaning' (19 participants), followed by `Model requirements', `Data collection', `Model training', `Model evaluation', and `Model deployment' activities, each having 17 participants out of 22. 5 out of 22 participants chose the `Other' option and elaborated on activities they engaged in through open-ended answers— activities that were not initially listed in the pre-interview questionnaire. Some of the mentioned activities included `system design', `data pipelines', `business benefit monitoring and reporting', `model integration developed by the research team into the pipelines', and `pipeline deployment'.

\subsection{RQ 1- What do AI practitioners \emph{understand} by ‘fair AI/ML’?} \label{sec:Definition}
\begin{figure}[htpb]
    \centering
    \includegraphics[width=\textwidth] {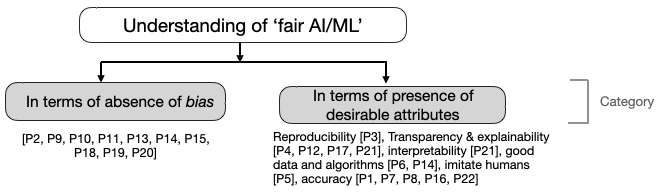}
    \caption{Overview of the participant's understanding of `fair AI/ML'}
    \label{fig:Definition}
\end{figure}

Based on the responses, we grouped the participants' understanding of `fair AI/ML' into two categories including, (i) In terms of absence of \emph{bias} and (ii) In terms of presence of desirable \emph{attributes}, which are explained in detail below. Figure \ref{fig:Definition} shows the overview of the participant's understanding of `fair AI/ML'. 

\subsubsection{In terms of absence of bias}
When a person is tasked with comprehending the concept of `fairness', it is quite probable that their understanding will revolve around the absence of biases as defined by tech companies such as Google \citep{Google}, IBM \citep{IBM}, and countries like Australia \citep{Australisethics}. When the participants were asked to share their understanding of `fair AI/ML', [P2, P9, P10, P11, P13, P14, P15, P18, P19, and P20]  described `fair AI/ML' in terms of absence of \emph{bias} in the AI/ML system. For example, participants [P9] and [P11] said:
\begin{quote}
\small
\faCommenting \hspace{0.05cm} \emph{``A fair model is a model which is not skewed and not biased."} - [P9]\\
\faCommenting \hspace{0.05cm} \emph{``So in my opinion, I guess like a fair model should be something which decreases the bias, as I mean, there should be very less bias."} - [P11]  
\end{quote}

 \subsubsection{In terms of the presence of desirable attributes:} 
[P1, P3, P4, P5, P6, P7, P8, P12, P14, P16, P17, P21 and P22] described `fair AI/ML' in terms of its features or attributes. The participants framed it as the necessary elements an AI/ML system must possess to be considered as \emph{fair}. For example, the participants said that the AI/ML system should be reproducible [P3], transparent and explainable [P4, P12, P17, P21], interpretable [P21], and accurate [P1, P7, P8, P16, P22]. Some participants also mentioned that a fair AI/ML system should use a good amount of data [P6] and have proper algorithms [P14].  For example, [P3], [P4], [P7], and [P14] said,
\begin{quote}
\small
\faCommenting \hspace{0.05cm} \emph{``But to me fairness is more about whether or not it is reproducible. It's something that can be tested and checked and improved from that again."} - [P3] \\
\faCommenting \hspace{0.05cm} \emph{``There should be transparency in any fair model that you build, right? So it should explain why it throws a certain outcome."} - [P4] \\
\faCommenting \hspace{0.05cm} \emph{``Fair model my understanding that could work with different data. It should still like give proper accurate results. There shouldn't be a huge difference between the seen data and unseen data."} - [P7]\\
\faCommenting \hspace{0.05cm} \emph{``A fair model should have proper algorithms which support you to treat your groups fairly and maybe post-processing stage where you when you're applying business logic."} - [P14] \\

\end{quote}

\subsection{RQ 2- What \emph{challenges} do AI practitioners face in developing a fair AI/ML system and what are the \emph{factors} that lead to those challenges?} \label{sec:Challenge}

\begin{figure}[htpb]
    \centering
    \includegraphics[width=\textwidth] {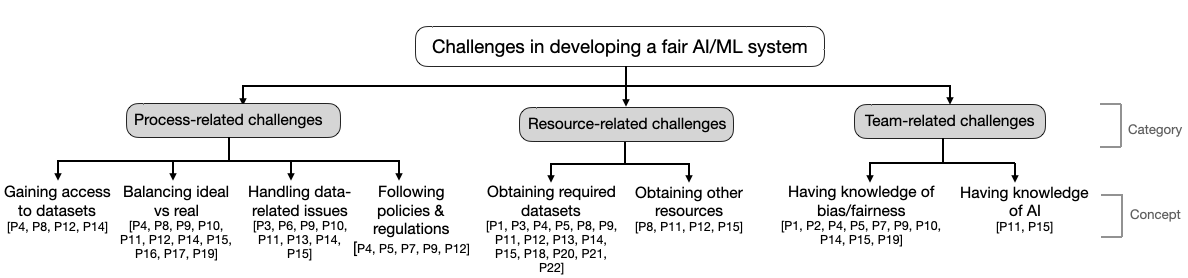}
    \caption{Overview of the participants' challenges in developing a fair AI/ML system}
    \label{fig:Challenges}
\end{figure}

We also asked the participants about the challenges they face in developing a fair AI/ML system through an open-ended question. We categorised the challenges of AI practitioners into three categories which are, (i) Process-related challenges, (ii) Resource-related challenges, and (iii) Team-related challenges, based on the responses of the participants. Each category is underpinned by multiple concepts and codes which are explained in detail below. Figure \ref{fig:Challenges} shows the overview of the challenges faced by AI practitioners in developing a fair AI/ML system. 

Once we inquired with the participants regarding the challenges they encountered in developing a fair AI/ML system, we delved further to understand the \emph{factors} leading to those challenges. Gaining insights into the factors leading to the challenges could contribute to devising more effective strategies to assist AI practitioners in overcoming those challenges. Additionally, we have highlighted the \emph{factors} leading to each challenge within the quotes of the participants. 

\subsubsection{Process-related challenges} \label{sec:process-related challenge}
The participants shared the challenges they encountered in developing fair AI/ML systems, specifically about the \emph{process} of developing such systems. Here, the term `process-related challenges' refers to the challenges faced during the development phase of the AI/ML systems. The participants reported four key challenges (concepts) under this category which include, (i) Gaining access to datasets, (ii) Balancing ideal vs real, (iii) Handling data-related issues, and (iv) Following policies and regulations. Each of these concepts is underpinned by multiple codes which are discussed below. 
\subsubsection*{\textbf{Gaining access to datasets}} 
AI practitioners involved in the development of AI/ML systems might face limitations in accessing vital resources for their work. Factors like adherence to company rules and regulations can impede their access to necessary resources, leading to challenges that, in turn, may contribute to the development of an unfair AI/ML system. In our study, [P4, P8, P12, and P14] reported the challenge of gaining access to the datasets they require to train an AI/ML model. The factors that led to the challenge of gaining access to the datasets include the size of the organisation and the data confidentiality policy. For example, participants [P4] and [P8] said,
\begin{quote}
   \small 

\faCommenting \hspace{0.05cm} \emph{``Like as a data scientist, I would have access to a very certain amount of data which I can pick. For example, some of the data would be from the external side or something you would not have access to that team data. Sometimes when you work due to \textbf{data confidentiality} policy, you will not have access to most of your data points."} - [P4] \\
\faCommenting \hspace{0.05cm} \emph{``Data is the one thing which models are built on but they are not available for public access, right? Like the open AI is things.. that language models are trained on the data sets, but data that are not available for \textbf{our cases (small companies)}."} - [P8] 
\end{quote}

\subsubsection*{\textbf{Balancing ideal vs real}}
\#Real-world data vs training data: The desire to develop a perfect AI/ML system is different from the ability to develop it. Training an AI/ML model with an extensive array of real-world data can be impractical. AI practitioners must rely on initial training data, and the system subsequently interacts with real-world data after deployment. Consequently, achieving fairness in AI/ML systems requires AI practitioners to navigate a balance between the training data and real-world data. This equilibrium can prove challenging at times due to a variety of limiting factors. In our study, [P8, P9, P10, P12, P15, P16, and P19] reported the challenge of striking a balance between the real-world data and the training data that they use in the development of AI/ML systems. The factors that led to the challenge of striking a balance between the real-world data and the training datasets include gaps existing between the real-world data and training data and negligence of the AI practitioners. For example, participants [P10] and [P15] said,
\begin{quote}
    \small

 \faCommenting \hspace{0.05cm} \emph{``Because the collected data is only a \textbf{subset of the data in real world}. So, the distribution of collected data is not identical to the real-world distribution, even if we do some data augmentation such as oversampling or other generative techniques, we cannot ensure the data distribution of augmented data is identical to the distribution in the real world. So we can just assume it is approximately identical, but they are not perfectly, identical. So it can be still a huge, huge challenge."} - [P10]\\
\faCommenting \hspace{0.05cm} \emph{``Originally, in the project we made, we were trying to do some stuff on human photos. And so we had to augment our data, but the way that the \textbf{data was created was not correct}, it didn't actually match the real data, like real photos. And so the AI actually learned to tell the difference between the synthetic data and the real data and could tell the difference between a fake photo and a real photo. And so that was like a kind of an eye opener for us that there was a bias from introduced from the synthetic data."} - [P15] \\
\end{quote}
\#Fairness requirements vs technical constraints: Developing an AI/ML system may appear straightforward, but it is a highly intricate undertaking. AI practitioners may face considerable challenges when trying to balance their envisioned ideal AI/ML system with the challenges of the real world. In our study, [P4, P11, P14, P16, P17, and P19] mentioned the challenge they face in maintaining a balance between their requirements for developing a fair AI/ML system with the technical constraints they encounter. They discussed factors such as the complex nature of AI and lack of time as contributing to this challenge. For example, [P16] and [P19] said:
\begin{quote}
    \small
\faCommenting \hspace{0.05cm} \emph{``And actually, at first, we do not talk about the machine learning bias. At first, we talk about the production because we are practitioners and it is more important, I mean, the \textbf{working version is the most important}."} - [P16]\\
\faCommenting \hspace{0.05 cm} \emph{``So as I mentioned, we tend to have like the model that works first, then we'll look at the virus later. In the industry here, every project has a \textbf{deadline and lifetime}. So if we don't launch the products, the stakeholders might not be happy and then they could find some other people who can do that."} - [P19]
\end{quote}

\subsubsection*{\textbf{Handling data-related issues}} \#Detecting data bias: One of the main aspects of developing any AI/ML system is data, presenting a significant challenge for AI practitioners in addressing data-related issues during the development of fair AI/ML system. Effectively managing biases in the data necessitates the initial step of detection, and corrective measures can only be taken once the biases are identified. The detection of data bias may pose a formidable challenge for AI practitioners, influenced by various factors. In our study, [P3, P6, P9, P10, P13, P14, and P15] reported the challenge of detecting data bias during the development of a fair AI/ML system. This challenge was led by factors like the nature of data bias, lack of time, and lack of tools/techniques. For example, [P9] and [P13] said:
\begin{quote}
   \small 
\faCommenting \hspace{0.05cm} \emph{``You're working on a \textbf{limited timeline project}. So your priority is to have a working model. And sometimes you might not be able to discover such data biases."} - [P9]\\
\faCommenting \hspace{0.05cm} \emph{If we have \textbf{data bias checking tool} to detect biases automatically, it would be great."} - [P13]
\end{quote}

\#Addressing data bias: Simply identifying data bias is not sufficient for developing a fair AI/ML system; it is crucial to actively address and rectify these biases. Dealing with identified bias in the data becomes the subsequent step and it can be challenging for AI practitioners due to several factors. In our study, [P9, P11, P14, and P15] discussed that they feel it is challenging to address (mitigate and/or remove) biases from the data during the development of AI/ML systems. The factors leading to the challenge of addressing data bias include a lack of tools/techniques and the biased nature of team members. For example, [P11] said,
\begin{quote}
\small
\faCommenting \hspace{0.05cm} \emph{``\textbf{If there was some kind of tool} which can let the person who is training the model know, maybe you need to remove this data, or else maybe you need to do these kinds of operations on your data, or maybe you need to do something, obviously, anyone wants that kind of tool as removing bias from the data is hard."} - [P11]
\end{quote}

\subsubsection*{\textbf{Following policies and regulations}}
 AI practitioners are required to adhere to various policies and regulations, which serve as guiding principles in developing AI/ML systems. Nevertheless, if these policies and regulations fall short, they can pose challenges to AI practitioners and can impede the development of a fair AI/ML system. Likewise, in our study, [P4, P5, P7, P9, and P12] expressed challenges in adhering to policies and regulations concerning AI ethics while working on the development of AI/ML systems. This challenge arose due to factors like lack of policies and lack of implementation of the policies. For example, [P5] and [P9] quoted:
 \begin{quote}
    \small 
\faCommenting \hspace{0.05cm} \emph{``I don't think Australia has any \textbf{updated AI ethics policies} and stuff, maybe, they need to update the policies based on how to follow the current trend and to follow the current technologies, something like that."} - [P5] \\
\faCommenting \hspace{0.05cm} \emph{``The European Alliance introduced a responsible AI framework but \textbf{not sure if Australia has adapted} such things in organisations. I know that companies like Facebook, are currently adapting. When it comes to small companies, small organisations, I'm not quite sure."} - [P9]
 \end{quote}

\subsubsection{Resource-related challenges} \label{sec:Resource-related challenge}
Participants encountering challenges in developing a fair AI/ML system face issues primarily associated with the \emph{resources} used in the development process, constituting the second category of challenges. The participants reported two key challenges (concepts) under this category which include, (i) Obtaining required datasets and (ii) Obtaining other resources. Each of these concepts is underpinned by multiple codes which are discussed below. 
\subsubsection*{\textbf{Obtaining required datasets}} 
As previously mentioned, data plays a crucial role in the development of AI/ML systems. In our study, the majority of the participants [P1, P3, P4, P5, P8, P9, P11, P13, P14, P15, P18, P20, P21, and P22] pointed out that obtaining the required datasets to develop a fair AI/ML system is challenging. The participants noted that they often acquire imperfect (non-representative) datasets or incomplete datasets for developing AI/ML systems, presenting a challenge in ensuring the fairness of such systems. The factors leading to this challenge include a lack of representative datasets, lack of cost, lack of tools/techniques, lack of control over data collection, and negligence of AI practitioners. For example, [P3], [P11] and [P20] said:
\begin{quote}
    \small
\faCommenting \hspace{0.05cm} \emph{``I think bias could happen since the first step because sometimes we have \textbf{other teams to collect the data}, yes, we don't know what they actually provide to us."} - [P3]\\
\faCommenting \hspace{0.05cm} \emph{``It has got something to do with the data, but then from the very beginning, like black people are discriminated in the world. So, I feel like the reason might be because, in the world, we have \textbf{more white people images than black people images}. So the data itself is less and it is a proven thing."} - [P11] \\
\faCommenting \hspace{0.05 cm} \emph{``Should I buy this data? Or not? It's money. In most cases, you don't use data for free. You know, sometimes you need to actually \textbf{pay for this in some way}."} - [P20]
\end{quote}

\subsubsection*{\textbf{Obtaining other resources}} 
\#Technological requirements: In addition to datasets, AI practitioners have diverse technological needs, including high-quality hardware, to augment the development of a fair AI/ML system. Technology plays a vital role in assisting AI practitioners in the development process. However, acquiring the necessary technology can present a challenge, potentially impeding the development of a fair AI/ML system. Only [P8, P11, and P12] indicated that they encounter challenges in developing fair AI/ML systems because they lack the necessary technology needed for the development process. The participants discussed the lack of cost as a factor leading to this challenge. For example, [P11] said:
\begin{quote}
  \small  
\faCommenting \hspace{0.05cm} \emph{``And resources because some huge models require huge GPUs. So in our company, we do not have GPUs because they are \textbf{expensive}. So yeah, lack of such resources of course is a challenge."} - [P11]
\end{quote}

\#Human-related requirements: Developing any software is a collaborative effort, involving multiple teams within a company dedicated to specific tasks. Collaborating with various team members offers advantages, such as diverse assistance in different aspects. Nonetheless, not all companies may incorporate multiple members in their development teams, posing a challenge for AI practitioners striving to develop a fair AI/ML system. In our study, [P12 and P15] reported that not having multiple people on the team is a challenge for them in developing a fair AI/ML system, and the lack of cost is the factor contributing to this challenge. For example, [P15] said:
\begin{quote}
   \small 
\faCommenting \hspace{0.05cm} \emph{``So even an individual (AI practitioner) can introduce bias into their model. And what we try to do is have the evaluation code be performed by a different person than the trainer. But, usually, \textbf{companies can't really afford} to do that which is challenging."} - [P15] 
\end{quote}

\subsubsection{Team-related challenges} \label{sec: team-related challenge}
The majority of participants also encountered challenges pertaining to their knowledge and understanding of different aspects, which we have classified as \emph{team}-related challenges in developing a fair AI/ML system. Here, the term `team-related challenges' refers to the challenges AI practitioners encounter as a result of their own limitations or shortcomings. The participants discussed two key challenges under this category which are the challenges in (i) Having knowledge of bias/fairness and (ii) Having knowledge of AI. These two key concepts are underpinned by multiple codes which are explained in detail below:

\subsubsection*{\textbf{Having knowledge of bias/fairness}} It is important that AI practitioners possess good knowledge and understanding of key concepts such as `bias' and `fairness' when aiming to develop a fair AI/ML system. However, AI practitioners might face difficulties in grasping the concepts of `bias' and `fairness' due to reasons such as the subjective nature of these concepts. In our study, [P1, P2, P4, P5, P7, P9, P10, P14, P15, and P19] reported that they are unable to understand the concept of `bias' or `fairness' which poses a challenge in developing a fair AI/ML system. The factors discussed by the participants that contributed to this challenge include a lack of domain knowledge, a lack of AI practitioners' common approach, and a lack of awareness. For example, [P4], [P9] and [P19] said, 
\begin{quote}
    \small
\faCommenting \hspace{0.05cm} \emph{``And again, now all humans' thoughts and the way \textbf{they approach a problem would be different}. So that is one more reason for not understanding the bias problems in the systems we develop."} - [P4] \\
\faCommenting \hspace{0.05cm} \emph{``Most of our data is from sensors. So we might not have like a very \textbf{clear view of biases} like the people who deal with NLP and stuff."} - [P9]\\
\faCommenting \hspace{0.05cm} \emph{``It's hard to understand fairness. I think I'm not intellectual enough to be in a position to define fairness. \textbf{My definition of fairness could be different from other people's} point of view."} - [P19]
\end{quote}

\subsubsection*{\textbf{Having knowledge of AI}} The rapid growth of AI is making it increasingly challenging for everyone to keep updated with its advancements and understand its outcomes \citep{pant2023ethics}. AI practitioners might face such challenges that can negatively impact the development of a fair AI/ML system. In our study, only [P11 and P15] reported that understanding AI outcomes is challenging due to its complex nature, negatively affecting the development of fair systems. For example, [P15] said, 
\begin{quote}
    \small
\faCommenting \hspace{0.05cm} \emph{``And then AI models, sometimes you don't know what it is actually deciding on and what it is actually measuring as it is too \textbf{complex}. So we had other cases where the AI kind of learns funny things that you don't anticipate."} - [P15]
\end{quote}

\subsection{RQ 3- What do AI practitioners perceive as the \emph{consequences} of developing an unfair AI/ML system?} \label{sec:consequence}
\begin{figure}[htpb]
    \centering
    \includegraphics[width=\textwidth] {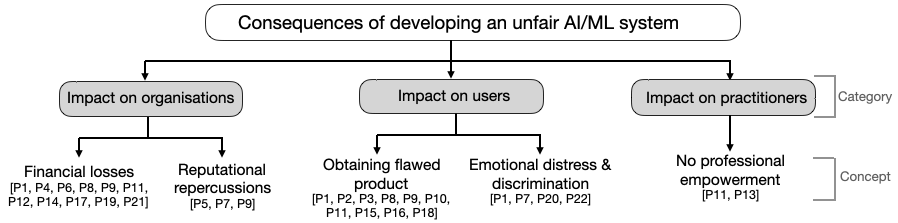}
    \caption{Overview of the consequences of developing an unfair AI/ML system}
    \label{fig:Consequence}
\end{figure}
We posed an open-ended query to the participants regarding the consequences of developing an unfair AI/ML system. The participants explored three distinct categories of negative consequences: (i) Impact on organisations, (ii) Impact on users, and (iii) Impact on practitioners. Each of these three categories is underpinned by multiple concepts and codes explained in detail below. Figure \ref{fig:Consequence} shows an overview of the consequences of developing an unfair AI/ML system. 

\subsubsection{Impact on organisations} The majority of participants delved into the adverse impacts on organisations resulting from the failure to develop a fair AI/ML system. The participants delved into two key facets (concepts) of the impacts on organisations namely: (i) Financial losses and (ii) Reputational repercussions.
\subsubsection*{\textbf{Financial losses}} Developing AI/ML systems is a complex process that demands various resources like time and money \citep{pant2023ethics}. If the systems turn out to be unfair or fail to meet goals, it not only affects the project but also leads to financial setbacks for the organisation. Many participants [P1, P4, P6, P8, P9, P11, P12, P14, P17, P19, and P21] in our study reported that the development of an unfair AI/ML system leads to financial losses to organisations. For example, participants [P4] and [P6] said:
\begin{quote}
    \small
\faCommenting \hspace{0.05cm} \emph{``Because the one that you're going to deploy would definitely have an impact on your business and in such cases, any small bias can lead to a huge financial loss."} - [P4] \\
\faCommenting \hspace{0.05cm} \emph{``Yeah, it also constitutes money loss to the organisation."} - [P6]
\end{quote}

\subsubsection*{\textbf{Reputational repercussions}} In today's tech-driven era, organisations are in fierce competition to enhance their software systems \citep{hua2020ai}. The constant race for improvement means even a minor flaw can tarnish an organisation's reputation. Developing an unfair AI/ML system poses a significant risk, as it can lead to severe reputational repercussions for organisations in this highly competitive landscape. In our study, only [P5, P7, and P9] provided insights into the consequences associated with the organisation's reputation when an unfair AI/ML system is developed. For example, participants [P5] and [P9] said:
\begin{quote}
    \small
\faCommenting \hspace{0.05cm} \emph{``If they are collecting the data, then there shouldn't be, any bias in the data. If bias is there, different kinds of controversies will rise in, and legal issues will arise."} - [P5]\\
\faCommenting \hspace{0.05cm} \emph{``But to give you another perspective, like Twitter, such models, are exposed to a large number of people and a large number of datasets and are heavily used worldwide. Such organisations need to ensure they do not have such biases in their datasets, or in finally at their models, because then it would create controversies, and then it could finally tarnish the image of these organisations as well."} - [P9]
\end{quote}

\subsubsection{Impact on users} \label{sec:Consequence_User} Many participants mentioned the negative impacts on users due to the development of an unfair AI/ML system. The participants discussed two key aspects (concepts) of the impacts on people including (i) Obtaining flawed product and (ii) Emotional distress and discrimination.

\subsubsection*{\textbf{Obtaining flawed product}} 
The primary objective of developing any AI/ML system is to aid users in various domains, be it healthcare, technology, education, etc. When AI/ML systems are developed unfairly, users receive flawed products, which undermines their core purpose. In our study, [P1, P2, P3, P8, P9, P10, P11, P15, P16, and P18] emphasised that the primary detriment to users resulting from the development of an unfair AI/ML system is the receipt of defective products, leading to inaccurate predictions. For example, [P1] and [P2] said:
\begin{quote}
    \small
\faCommenting \hspace{0.05cm} \emph{``It is important to create a fair ML model because the main reason is that if the ML model is biased, users won't be able to achieve our end goal. If the model is biased, it won't give the result that it is required to give."} - [P1] \\
\faCommenting \hspace{0.05cm} \emph{``Users will be getting a great product from the company if it is fair, otherwise not."} - [P2]
\end{quote}

\subsubsection*{\textbf{Emotional distress and discrimination}} When an unfair AI/ML system is developed, most likely the users who are using those systems are impacted negatively \citep{Amazon, prates2020assessing}. The primary impact could manifest as emotional distress and discrimination towards these users. Only [P1, P7, P20, and P22] highlighted that users experience emotional distress and discrimination when using unfair AI/ML systems, leading to hurt sentiments. For example, a participant [P1] said:
\begin{quote}
    \small
\faCommenting \hspace{0.05cm} \emph{``The second reason is that the sentiments of the people can be hurt if the model is biased."} - [P1]
\end{quote}

\subsubsection{Impact on practitioners} A very few participants mentioned the negative impacts of developing an unfair AI/ML system on the practitioners responsible for their development. We classified these negative consequences into a specific aspect: (i) No professional empowerment. 
\subsubsection*{\textbf{No professional empowerment}}
AI practitioners gain valuable insights through hands-on experience in developing AI/ML systems, complementing their theoretical knowledge. The development of unfair systems, however, could have a negative impact on these practitioners, affecting their learning experiences in the field. Only [P11 and P13] highlighted that the development of unfair AI/ML systems hinders the professional empowerment of AI practitioners, causing a decline in confidence and knowledge. For example, participants [P11] and [P13] said, 
\begin{quote}
    \small
\faCommenting \hspace{0.05cm} \emph{``When they develop an unfair model, they will not learn something new from that.. like, in terms of data augmentation, or terms of algorithmic change, or terms of data collection."} - [P11]\\
\faCommenting \hspace{0.05cm} \emph{``Yeah, if the research team or the model training team make the unfair model and we lack the confidence to use the model directly in our product."} - [P13]
\end{quote}

\subsection{RQ 4- What \emph{strategies} do AI practitioners use in ensuring the fairness of an AI/ML system?} \label{sec:Strategy}

\begin{figure}[htpb]
    \centering
    \includegraphics[width=\textwidth] {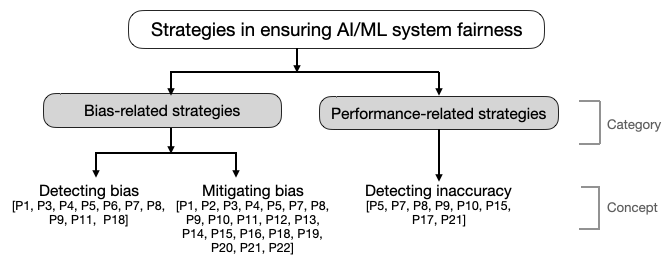}
    \caption{Overview of the participants' strategies in ensuring the fairness of an AI/ML system}
    \label{fig:Strategy}
\end{figure}
We asked the participants about the strategies that they use to ensure the fairness of AI/ML systems. The participants discussed two categories of strategies that they use to ensure the fairness of AI/ML systems which are (i) Bias-related strategies and (ii) Performance-related strategies. Each of these two categories is underpinned by multiple concepts and codes explained in detail below. Figure \ref{fig:Strategy} shows the overview of AI practitioners' strategies to ensure the fairness of the AI/ML systems they develop.

\subsubsection{Bias-related strategies} \label{sec:bias-related strategy}
The majority of the participants discussed the strategies they used to address the bias-related issues when developing a fair AI/ML system. The participants reported two key strategies (concepts) they used to address bias-related issues including (i) Detecting bias and (ii) Mitigating bias. Each of these concepts is discussed in detail below. 

\subsubsection*{\textbf{Detecting bias}} Addressing any concern starts with its detection; without identification, mitigation is impossible. In our study also, participants outlined their strategies for ensuring the fairness of the AI/ML systems they developed, emphasising the initial step of detecting bias. [P1, P3, P4, P5, P6, P7, P8, P9, P11, and P18] mentioned that they rely on testing the system with test datasets as their strategy for identifying biases in the system. For example, participants [P1] and [P8] said:
\begin{quote}
    \small
\faCommenting \hspace{0.05cm} \emph{`` So in the beginning, we categorise our whole dataset as train, test, and validation dataset and use the testing data to test the model. The testing phase is mandatory, otherwise, we won't know if the model we created has biases."} - [P1]\\
\faCommenting \hspace{0.05cm} \emph{``After building the model, we use some test case scenarios and we have been doing this post process like how well the model is doing in the test data set to find out any biases."} - [P8]
\end{quote}

\subsubsection*{\textbf{Mitigating bias}} 
\#By balancing datasets: Data holds significance in the development of AI/ML systems. Using well-balanced datasets for training is important to ensure the system generates an unbiased prediction. Likewise, in our study, the majority of the participants [P1, P3, P5, P7, P8, P9, P10, P11, P13, P14, P15, P16, P18, P19, P20, P21, and P22] mentioned using the data augmentation technique to balance training datasets during the development phase to mitigate biases in their systems. For example, participants [P14] and [P22] said:
\begin{quote}
    \small

\faCommenting \hspace{0.05cm} \emph{``I try to identify whether there are any segments where these metrics are up or down, and then I would go back to the data and see if it is because we don't have enough data for those regions. And then if that is the case, I'll try to maybe augment the data."} - [P14]\\
\faCommenting \hspace{0.05cm} \emph{I tried to increase for those that did not have enough by strategically doing some kind of artificial, like stretching the existing data or compressing data because I was working with audio data."} - [P22] 
\end{quote}

\#By involving multiple people: Collaborating in a team with multiple members offers numerous advantages in software development, fostering the exchange of knowledge and facilitating mutual learning \citep{augustin2002accelerating}. This collaborative dynamic can also prove particularly beneficial in the context of developing AI/ML systems. In our study, [P2, P7, and P12] mentioned that they get input from multiple people and this collaborative approach aids them in mitigating data biases within the system they develop. For example, [P7] said:
\begin{quote}
    \small
\faCommenting \hspace{0.05cm} \emph{``I don't have any medical background. Sometimes, some factors or features I never thought about could cause bias. So we have other advisors, like from another university, doctors, and professors. So yeah, we have that on the system to ask them (domain experts) if we have any doubts."} - [P7]
\end{quote}

\#By focusing on practices: In addition to securing the necessary resources, adhering to best practices can be important in the development of an AI/ML system. In our study, [P1, P2, P4, P5, and P12] reported that focusing on the practices of developing a fair AI/ML system is the strategy they take in mitigating biases from the system. For example, [P2] and [P4] quoted:
\begin{quote}
    \small
\faCommenting \hspace{0.05cm} \emph{``I just become conscious about unconscious biases during the development process."} - [P2]\\
\faCommenting \hspace{0.05cm} \emph{``I would also believe in the feedback mechanism out there, not just seeing your results on the test set and then going and deploying it, but rather enabling a feedback mechanism. And whenever the system goes a little off in terms of prediction, immediately, the feedback loop is getting connected there. So that is one way that I generally rectify my bias."} - [P4] 
\end{quote}

\subsubsection{Performance-related strategies}
A few participants elaborated on strategies for the performance of the AI/ML system they developed to ensure fairness. Within this category, participants deliberated on a specific strategy (concept), namely, (i) Detecting inaccuracy, which is explained below.

\subsubsection*{\textbf{Detecting inaccuracy}} 
In our study, [P5, P7, P8, P9, P10, P15, P17, and P21] mentioned that they detected the inaccuracy of the AI/ML system by using evaluation metrics. This helps them gauge the system's performance and ensure its fairness. For example, [P8] said:
\begin{quote}
    \small
\faCommenting \hspace{0.05cm} \emph{``Fair model should be 100\% accurate. 100\% accuracy is good for our scenario, but we also have other case scenarios like the loss and other things to notice in the evaluation metrics. When we have to check what is the loss of the models we have to go for the minimum loss."} - [P8] \\

\end{quote}
\subsection{Summary of Key Findings}
This study focuses on exploring AI practitioners' \emph{understanding} of `fair AI/ML', exploring the \emph{challenges} they encounter during the development of a fair AI/ML system, understanding the \emph{consequences} of developing an unfair AI/ML system perceived by them and investigating the \emph{strategies} they employ to ensure fairness in the AI/ML systems they develop. Table \ref{table:KF} shows the summary of the key findings of our study. 

\begin{table}[htbp]
\centering
\caption{Key Findings (KF) of the study.} \label{table:KF}     
\scriptsize
\begin{tabular} {>{\raggedright\arraybackslash}p{0.5cm}>{\raggedright\arraybackslash}p{10.5cm}>{\raggedright\arraybackslash}p{0.6cm}}

\hline\noalign{\smallskip}
 & Key Findings (KF) & Section\\

\hline\noalign{\smallskip}
KF1 & \textbf{AI practitioners' understanding of `fair AI/ML'}  & \ref{sec:Definition}  \\

 & (i) In terms of the absence of \emph{bias} &     \\

 & (ii) In terms of the presence of desirable \emph{attributes} (transparency, accuracy, interpretability etc.) &    \\

KF2 &  \textbf{AI practitioners' challenges in developing a fair AI/ML system} & \ref{sec:Challenge}  \\

 & (i) Process-related challenges: gaining access to datasets, balancing ideal vs real, handling data-related issues, and following policies and regulations &     \\

& (ii) Resource-related challenges: obtaining required datasets and obtaining other resources & \\

& (iii) Team-related challenges: having knowledge of bias/fairness and having knowledge of AI & \\

KF3 & \textbf{Consequences of developing an unfair AI/ML system perceived by AI practitioners}  & \ref{sec:consequence} \\

 & (i) Impact on organisations: financial losses and reputational repercussions &  \\
 & (ii) Impact on users: obtaining flawed product and emotional distress and discrimination & \\
 & (iii) Impact on practitioners: no professional empowerment & \\

KF4 & \textbf{AI practitioners' strategies to ensure the fairness of an AI/ML system} & \ref{sec:Strategy} \\

 & (i) Bias-related strategies: detecting bias and mitigating bias &  \\
& (ii) Performance-related strategies: detecting inaccuracy &\\

KF5 & Despite differing understandings of `fair AI/ML' among practitioners, a common challenge they report facing is obtaining the required datasets for model training during AI/ML system development. & \ref{def vs challenge} \\

KF6 & Some AI practitioners describing `fair AI/ML' in terms of absence of \emph{bias} seemed to have a broader understanding of negative consequences on practitioners due to the development of unfair AI/ML systems compared to those emphasising the presence of desirable \emph{attributes}. & \ref{def vs consequence} \\

KF7 & Despite differing understandings of `fair AI/ML' among practitioners, a common strategy they report using to ensure fairness in AI/ML systems is implementing bias-mitigation strategies. & \ref{def vs strategy} \\

\noalign{\smallskip}\hline
\end{tabular}
\end{table}

\subsection {Framework showing the relationship between the aspects-\emph{understanding}, \emph{challenges}, \emph{consequences} and \emph{strategies}}
\begin{figure*}[htpb]
    \centering
    \includegraphics[width=\textwidth] {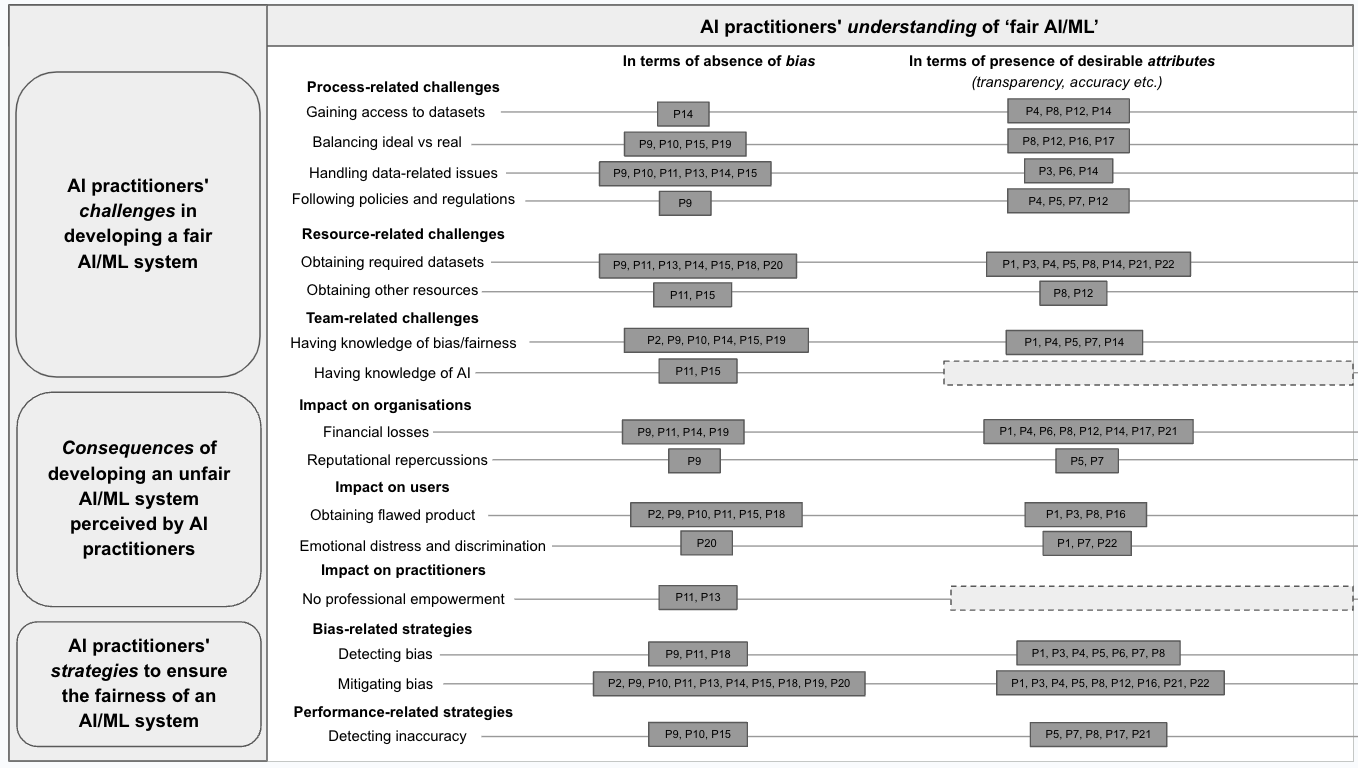}
    \caption{A framework showing the relationship between AI practitioners' \emph{understanding} of `fair AI/ML', with three aspects including, (i) their \emph{challenges} in its development, (ii) \emph{consequences} of developing unfair AI/ML system perceived by them, and (iii) their \emph{strategies} in ensuring fairness of AI/ML systems}
    \label{fig:Plot}
\end{figure*}

In this section, we discuss the relationship between AI practitioners' \emph{understanding} of `fair AI/ML' with three other aspects including, (i) the \emph{challenges} encountered in the development of a fair AI/ML system, (ii) the \emph{consequences} of developing an unfair AI/ML system, and (iii) \emph{strategies} used to ensure the fairness of an AI/ML system. Every participant [P1 to P22] in our study described `fair AI/ML' either in terms of the absence of \emph{bias} or in terms of the presence of desirable \emph{attributes} in AI/ML systems. The exception was the participant [P14], who described it in terms of the absence of \emph{bias}, as well as in terms of the presence of desirable \emph{attributes} in AI/ML systems. To illustrate the relationship between these aspects, we developed a framework, which is shown in Figure \ref{fig:Plot}. 

\subsubsection{Relationship between AI practitioners' \textbf{\emph{understanding}} and their \textbf{\emph{challenges}}} \label{def vs challenge}

\subsubsection*{\textbf{AI practitioners' \emph{understanding} -- in terms of the \emph{absence of bias} vs \emph{challenges}}}
The interview indicates that the participants who described `fair AI/ML' in terms of the absence of \emph{bias} in an AI/ML system reported challenges in all three categories: (i) process-related, (ii) resource-related, and (iii) team-related challenges that they face during AI/ML system development, as shown in Figure \ref{fig:Plot}. Specifically, the majority of the participants reported facing challenges in obtaining required datasets (resource-related challenge), followed by the challenge of handling data-related issues (process-related challenge). We see that AI practitioners who described `fair AI/ML' in terms of the absence of \emph{bias} in AI/ML systems also reported facing substantial challenges related to the datasets used in development. These challenges primarily revolve around resource availability, specifically in obtaining the necessary datasets and handling data-related issues.

\subsubsection*{\textbf{AI practitioners' \emph{understanding} -- in terms of the \emph{presence of desirable attributes} vs challenges}}
The interview shows that the participants who described `fair AI/ML' in terms of the presence of desirable \emph{attributes} reported almost all the challenges across all three categories— (i) process-related, (ii) resource-related, and (iii) team-related challenges that they encounter during the development of an AI/ML system, as illustrated in Figure \ref{fig:Plot}. Notably, none of the participants reported the challenge of lacking knowledge of AI (team-related challenge), whereas, most of the participants reported facing the challenge of obtaining required datasets (resource-related challenge) during AI/ML system development. 

In summary, regardless of how AI practitioners described `fair AI/ML', a common challenge they faced was obtaining the required datasets to train a model during AI/ML system development.

\subsubsection{Relationship between AI practitioners' \textbf{\emph{understanding}} and the \textbf{\emph{consequences}}}\label{def vs consequence}

\subsubsection*{\textbf{AI practitioners' \emph{understanding} -- in terms of the \emph{absence of bias} vs \emph{consequences}}}
The interview indicates that the participants in our study, describing `fair AI/ML' in terms of the absence of \emph{bias} in AI/ML systems, perceived the negative consequences of developing an unfair AI/ML system across all three categories including, (i) impact on organisations, (ii) impact of users and (iii) impact on practitioners, as shown in Figure \ref{fig:Plot}. Notably, the majority of participants perceived the acquisition of flawed products by users as a negative consequence of developing an unfair AI/ML system. However, the majority did not explicitly mention emotional distress and discrimination for users, as well as reputational repercussions for organisations, as significant negative consequences. These two specific concerns were expressed by only one participant each. 

\subsubsection*{\textbf{AI practitioners' \emph{understanding} -- in terms of the \emph{presence of desirable attributes} vs \emph{consequences}}}
The interview shows that the participants in our study, describing `fair AI/ML' in terms of the presence of desirable \emph{attributes} perceived the negative consequences of developing an unfair AI/ML system across only two categories including, (i) impact on organisations, and (ii) impact on users. According to the data, most participants talked about the financial loss to organisations as a negative consequence of developing an unfair AI/ML system, as shown in Figure \ref{fig:Plot}. Notably, no participants in our study mentioned any negative consequences of developing an unfair AI/ML system for the practitioners engaged in their development.

In summary, of the ones who described `fair AI/ML' in terms of absence of \emph{bias},  two of them discussed the negative consequence of developing an unfair AI/ML system on practitioners involved in the development. While those who described it in terms of the presence of desirable \emph{attributes} did not discuss the negative consequences of developing an unfair AI/ML system for practitioners at all. It looks like the former were able to acknowledge it, and they seem to have a broader understanding of the negative consequences associated with developing unfair AI/ML systems.   
   
  \subsubsection{Relationship between AI practitioners' \textbf{\emph{understanding}} and their \textbf{\emph{strategies}}} \label{def vs strategy}

\subsubsection*{\textbf{AI practitioners' \emph{understanding} -- in terms of the \emph{absence of bias} vs \emph{strategies}}}
According to the interview, the participants in our study who described `fair AI/ML' in terms of the absence of \emph{bias} in AI/ML systems discussed strategies falling into both categories, including (i) bias-related strategies and (ii) performance-related strategies. It shows that most participants discussed the strategies to mitigate bias (bias-related strategies) from AI/ML systems to ensure its fairness. In contrast, only a small number of participants reported the strategies of detecting bias (bias-related strategies) and detecting inaccuracy (performance-related strategies) to ensure the fairness of AI/ML systems. 

\subsubsection*{\textbf{AI practitioners' \emph{understanding} -- in terms of the \emph{presence of desirable attributes} vs \emph{strategies}}}
The interview shows that the participants in our study who described `fair AI/ML' in terms of the presence of desirable \emph{attributes} discussed strategies falling into both categories, including (i) bias-related strategies and (ii) performance -related strategies. It shows that a slightly higher number of participants discussed the strategy of mitigating bias (bias-related strategies) from AI/ML systems to ensure its fairness as compared to other strategies like detecting bias (bias-related strategies) and detecting inaccuracy (performance-related strategies). In contrast, an almost equal number of participants discussed strategies like detecting bias (bias-related strategies) or detecting inaccuracy (performance-related strategies) to ensure AI/ML system fairness. 

In summary, the interview shows a consistent trend among participants discussing the strategies that they used to ensure the fairness of AI/ML systems, regardless of how they described `fair AI/ML'. The majority of the participants who described `fair AI/ML' either in terms of the absence of \emph{bias} as well as in terms of the presence of desirable \emph{attributes} in AI/ML systems reported the use of the common strategy of mitigating bias (bias-related strategies) to ensure the fairness of the AI/ML systems they developed. 

\section{Discussion} \label{sec:Discussion}
In this section, we discuss and compare our findings in light of the related works.

\subsection{Definition/understanding of AI/ML fairness} \label{sec:Def_Discussion}
In recent years, major players in the tech industry, such as Google, Microsoft, and IBM, have delved deeply into the concept of \emph{fairness} in AI. Their consensus on the `fairness' principle revolves around minimising bias and fostering inclusive representation in the development of AI \citep{Google, Microsoft, IBM}. 

Various experiments, including those by \citet{harrison2020empirical} and \citet{srivastava2019mathematical}, have explored user perspectives on AI/ML fairness. The study conducted by \citet{harrison2020empirical} with non-technical users on Amazon Mechanical Turk (AMT) revealed that users reported unbiased models might not be automatically perceived as fair. This finding does not align with our study as some participants described `fair AI/ML' in terms of absence of \emph{bias}. On the other hand, the experiment by \citet{srivastava2019mathematical} on AMT found users defining fairness technically, focusing on accuracy and demographic parity, mirroring our study where AI practitioners also described `fair AI/ML' in terms of accuracy (the presence of desirable \emph{attributes}) of the system. Importantly, our study involved AI practitioners, while the mentioned studies focused on the general users of AI/ML systems.

Due to the lack of research that focuses on investigating AI practitioners' understanding of `fair AI/ML', we conducted an empirical study with 22 AI practitioners to investigate their understanding of what a `fair AI/ML' is. In our study, AI practitioners described `fair AI/ML' in terms of the absence of \emph{bias} and in terms of the presence of desirable \emph{attributes} in AI/ML systems. \citet{ryan2023integrating} in their empirical study, found that participants when discussing the term `fairness', commonly focused on preventing biased decisions of ML systems. This aligns with our findings as several AI practitioners in our study also described `fair AI/ML' in terms of the absence of \emph{bias}. It is important to note that both academic and industry professionals in the fields of Human-Computer Interaction (HCI) and ML participated in \citet{ryan2023integrating}'s study. Similarly, aligning with the definition of \emph{`fairness'} introduced by tech companies like Google \citep{Google} and Microsoft \citep{Microsoft}, some AI practitioners in our study described `fair AI/ML' in terms of the absence of \emph{bias}. However, when describing in terms of the presence of desirable \emph{attributes} in AI/ML systems, AI practitioners in our study specified features such as interpretability, transparency, and explainability that an AI/ML system should possess to be deemed \emph{fair}. \citet{ryan2023integrating} also highlighted that a few participants mentioned that a system needs to be transparent to be considered fair. Notably, principles like `explainability' and `transparency' are outlined separately in the AI ethics principles listed by tech companies such as Google \citep{Google}, IBM \citep{IBM}, Microsoft \citep{Microsoft}, and countries such as Australia \citep{Australisethics} and countries in Europe \citep{EUethics}. This suggests a lack of alignment between how AI practitioners understand `fair AI/ML' and the definitions set forth by tech companies, countries, and continents. This misalignment may hinder the development of universally accepted principles for fair AI/ML systems, potentially resulting in disparate approaches and interpretations within the AI community. 

In a similar vein, \emph{accuracy} and \emph{fairness} are categorised as two different non-functional requirements of an ML system \citep{habibullah2023non}. The \emph{accuracy} of an ML system has been categorised as a non-functional requirement, which can be measured using ML-specific or standard measures, whereas \emph{fairness} has been categorised as a non-functional requirement that cannot be measured and is non- quantifiable \citep{habibullah2023non}. However, in our study, we found that the participants [P1, P7, P8, P16, and P22] described `fair AI/ML' in terms of \emph{accuracy} of the AI/ML system (Section \ref{sec:Definition}). Along with that, when asked about the strategies to ensure AI/ML fairness, some participants [P5, P7, P8, P9, P10, P15, P17, and P21] reported that they focus on detecting the inaccuracy of the system (Section \ref{sec:Strategy}). The participants in our study considered \textit{accuracy} as a requirement to develop a fair AI/ML system. This also shows that the way AI practitioners in our study described `fair AI/ML' is different from the definitions of \emph{fairness} provided by tech companies like Google, IBM, etc, and different countries/continents like the USA, China, Australia, Europe, etc. Notably, these definitions do not include considerations regarding the \emph{accuracy} of AI/ML systems. Understanding how AI practitioners conceptualise `fair AI/ML' is important for developing effective policies and guidelines on \emph{AI fairness}. By incorporating their perspectives, frameworks can be created that not only meet regulatory standards but also align with practical implementation challenges and fairness considerations faced in real-world AI applications. This can lead to more fair, inclusive, and socially responsible AI/ML systems.

\subsection{Challenges in developing a fair AI/ML system}
Studies have highlighted challenges for AI practitioners in developing a fair AI/ML system across various domains and phases of development. Our study specifically focused on investigating the overall challenges of AI practitioners in developing a fair AI/ML system through semi-structured interviews. A majority of participants in the studies by \citet{holstein2019improving,fenu2022experts} faced challenges related to limited control over data collection, as well as challenges in obtaining balanced and representative datasets for model training due to a lack of methods supporting data collection and curation \citep{holstein2019improving}. These findings align with our study, where participants reported similar challenges in obtaining necessary datasets, attributing them to a lack of control over data collection, and expressed difficulties in obtaining balanced datasets due to a lack of methods for data collection and curation (Section \ref{sec:Resource-related challenge}). Most participants in \citet{holstein2019improving}'s study reported challenges in detecting biases in the ML system, due to a lack of support and challenges in developing their own solutions due to limited time. Our findings align with these, as participants in our study also highlighted how constraints, such as lack of support and time, pose challenges in detecting biases in systems and developing their envisioned ideal system (Section \ref{sec:process-related challenge}). 
Similarly, \citet{madaio2022assessing} identified collecting datasets as a challenge for AI practitioners during the development of AI systems, primarily due to the need to safeguard the personal information of user data. Our study's findings align, as participants also reported facing challenges in accessing necessary datasets due to data confidentiality concerns (Section \ref{sec:process-related challenge}). \citet{madaio2022assessing} found that participants reported challenges related to the resources required to develop a fair AI system, which aligns with one of our findings. Participants in our study also reported challenges in obtaining resources such as datasets, technology, and human resources, as discussed in Section \ref{sec:Resource-related challenge}. Both studies identified funding issues as a contributing factor to this challenge. \citet{fenu2022experts} reported that participants faced challenges in collecting data for training an AI system due to a lack of datasets representing the diversity of the population, which aligns with our findings (Section \ref{sec:Resource-related challenge}).
\citet{fenu2022experts} highlighted that adhering to regulations related to the fairness of an AI system was a reported challenge. Some participants in our study also emphasised challenges in following policies and regulations, citing reasons such as the lack of policies and inadequate implementation as discussed in Section \ref{sec:process-related challenge}. Likewise, \citet{hopkins2021machine} in their empirical study reported challenges faced by ML practitioners in detecting bias in ML, attributed to biased data or insufficient model testing, as discussed in Section \ref{sec:process-related challenge}. The participants in our study also emphasised the same challenge, citing factors such as the nature of data bias, lack of time, and a lack of tools/techniques to assist in detecting system biases. \citet{ryan2023integrating} in their empirical study highlighted challenges anticipated by HCI experts and ML experts in developing a fair AI which includes obtaining high-quality data to develop and evaluate a model, which aligns with our findings (Section \ref{sec:Resource-related challenge}). However, \citet{ryan2023integrating} identified additional challenges, including the importance of clarity regarding the model's context and credibility, as well as the difficulty in aligning the mathematical definition of fairness with the accuracy of the model, which does not align with our findings.

While some findings in our study align with previous research, there are unique contributions, particularly in uncovering team-related challenges faced by AI practitioners in developing a fair AI/ML system. Our study uncovered challenges specific to the development team, such as having knowledge of bias/fairness and knowledge of AI, as discussed in Section \ref{sec: team-related challenge}. Understanding these challenges is crucial, given that AI practitioners play a pivotal role in designing systems that have substantial societal impact and it fosters responsible and effective AI development \citep{orr2020attributions}. Similarly, our study uncovered challenges AI practitioners face in balancing real-world data with training data in AI/ML system development (Section \ref{sec:process-related challenge}), a finding not reported in previous studies. Additionally, we identified challenges related to obtaining various resources, including technological and human-related resources, which impact the development of a fair AI/ML system (Section \ref{sec:Resource-related challenge}). These insights contribute new dimensions to the existing understanding of challenges in this domain. Addressing these challenges of AI practitioners can help in developing fair AI/ML systems that can be crucial for mitigating societal inequalities and promoting fairness in society \citep{holstein2019improving}.  

\subsection{Consequences of developing an unfair AI/ML system}
Studies have explored the consequences of developing an unfair AI/ML system from the perspectives of different stakeholders \citep{marcinkowski2020implications, shin2019role}. For example, \citet{woodruff2018qualitative} identified that users reported the potential negative consequences of algorithmic unfairness which include racial discrimination and stereotyping and loss of opportunities for personal advancement.   

\citet{weidener2024role} found that AI experts reported fatal outcomes for users from unfair AI-based systems. Our findings differ, as our participants did not mention fatal outcomes but highlighted other impacts such as users obtaining flawed products, and facing emotional distress, and discrimination as described in Section \ref{sec:consequence}. Given the limited research on AI practitioners' perspectives on the consequences of developing an unfair AI/ML system, we conducted semi-structured interviews with 22 practitioners. Our study revealed new insights, identifying three main negative consequences perceived by AI practitioners: those affecting organisations, users, and the practitioners themselves, as discussed in Section \ref{sec:consequence}. Understanding the consequences of developing an unfair AI/ML system may facilitate the development of specific mitigation strategies. Addressing issues at the organisational, user, and practitioner levels may contribute to more effective and comprehensive solutions in tackling unfairness in AI/ML systems. For instance, our study revealed that only a small number of practitioners acknowledge the negative impact on users when developing unfair AI/ML systems (Section \ref{sec:Consequence_User}). This highlights a critical gap in considering user perspectives during AI/ML development, emphasising the need for more user-centric approaches \citep{dankloff2024analysing}. Developing such user-centric systems is essential for fostering user trust in AI/ML systems, ensuring fairness and reliability. 

\subsection{Strategies in ensuring fairness in AI/ML systems}
In recent years, numerous studies have been conducted exploring strategies and approaches to AI/ML fairness. Several qualitative studies, such as those by \citet{deng2022exploring}, \citet{richardson2021towards}, and \citet{balayn2023fairness}, have explored AI practitioners' experiences and perspectives on specific fairness toolkits. These studies conducted semi-structured interviews with AI/ML practitioners to understand their practices in using different fairness toolkits. However, as our study focuses on general strategies employed by AI practitioners to ensure fairness in AI/ML systems, the findings from these studies do not align with our study.

On the other hand, \citet{madaio2020co} identified AI practitioners' processes for recognising and addressing fairness issues in AI systems, emphasising understanding fairness as a personal priority and adhering to ad-hoc processes. However, these findings diverge from our study, which concentrates on tactical approaches or strategies used by AI practitioners in their day-to-day lives to ensure the fairness of AI/ML systems. \citet{ryan2023integrating} noted that the common approach used by ML and HCI experts when addressing fairness was associated with data used in an AI system. This finding aligns with our study as most of the participants in our study also discussed the bias mitigation strategy in the AI/ML system by balancing datasets (Section \ref{sec:bias-related strategy}). Similarly, in the study by \citet{ryan2023integrating}, a participant mentioned comparing model accuracy across demographic groups to assess fairness. This corresponds with our findings, where several participants also identified inaccuracies in AI/ML systems through the use of evaluation metrics. However, participants in \citet{ryan2023integrating}'s study mentioned not considering fairness in the AI systems they develop, contrasting with our findings. In our study, each participant reported employing at least one strategy to ensure fairness in AI/ML systems. 

Our study presents unique contributions, notably in revealing strategies employed by AI practitioners to detect bias for ensuring fairness in AI/ML systems, as discussed in Section \ref{sec:bias-related strategy}. These insights, including collaboration with team members to mitigate data biases and a focus on individual practices during AI/ML system development, are novel findings that have not been reported in previous research.

\subsection{Insights} \label{sec:Insights}
Based on the memos written for the study, we uncovered several interesting insights and reflections. Research recommendations can be made based on these findings and our insights.

\subsubsection{`No bias'- necessary but not sufficient to make a fair AI/ML system} 
Most ethical guidelines in AI stress the importance of ensuring \emph{fairness}, aiming to eliminate bias and discrimination within AI systems. For example, Australia's AI Ethics Principles defined \emph{`fairness'} as \emph{``AI systems should be inclusive and accessible, and should not involve or result in unfair discrimination against individuals, communities or groups"} \citep{Australisethics}. Similarly, the European Commission defined `Diversity, non-discrimination and fairness' as, \emph{``Unfair bias must be avoided, as it could have multiple negative implications, from the marginalisation of vulnerable groups to the exacerbation of prejudice and discrimination. Fostering diversity, AI systems should be accessible to all, regardless of any disability, and involve relevant stakeholders throughout their entire life circle"} \citep{EUethics}. However, based on the interview with AI practitioners, we identified that `no bias' is a crucial element in developing a fair AI/ML system, but it alone is inadequate. For example, when participants were asked to share their understanding of `fair AI/ML', they noted that it must not only be accurate but also exhibit attributes such as transparency and reproducibility to qualify as a fair AI/ML system, as discussed in Section \ref{sec:Definition}. Some research has also discussed similar notions. For example, \citet{silberg2019notes} reported that the absence of unwanted bias is insufficient to infer that the AI system is `fair'. Similarly, participants in an experiment perceived that an unbiased AI/ML system does not necessarily mean it is perceived as fair, as they often found certain systems unfair despite being unbiased, especially when errors were distributed unevenly among different racial groups \citep{harrison2020empirical}. 

Even though participants had broader ideas about what constitutes a `fair AI/ML', it was interesting to observe that their discussions predominantly revolved around the concept of `bias' when responding to various questions. For instance, when queried about \emph{strategies} employed to ensure the fairness of an AI/ML system, the majority focused on detecting or addressing biases in AI/ML systems. In contrast, only a small number delved into strategies related to optimising the system's performance for fairness, as outlined in Section \ref{sec:Strategy}. Similarly, in discussions about the \emph{consequences} of developing an unfair AI/ML system, nearly all participants used the term `bias' and elaborated on the consequences of developing a `biased' system. Hence, while achieving a `no bias' is crucial for developing a fair AI/ML system, it is essential to recognise that it alone is not adequate; nevertheless, it remains a significant aspect in the development of a fair AI/ML system.

\subsubsection{Data bias vs other biases in AI/ML systems}
As previously discussed, we observed that participants primarily centered their discussions on the notion of `bias' when working towards a fair AI/ML system. Machine learning (ML) can be prone to various biases, including biases from data to algorithm, algorithm to user, and user to data. Moreover, each of these categories encompasses different sub-types of biases \citep{mehrabi2021survey}. However, in our case, even within the discourse on `bias', a majority of participants specifically addressed the concept of `data bias'. For example, when the participants were asked about the challenges encountered in developing a fair AI/ML system, one of the aspects (concepts) they highlighted pertained to handling data-related issues, as detailed in Section \ref{sec:process-related challenge}. Likewise, the majority of the participants discussed the strategies related to detecting bias in the data and mitigating data biases when they were asked about the strategies to ensure the fairness of the AI/ML system as discussed in Section \ref{sec:bias-related strategy}. Additionally, participants discussed the factors leading to their challenges to develop a fair AI/ML system, mainly related to data bias, such as the use of biased training data, lack of tools to check data bias, etc. as discussed in Section \ref{sec:Challenge}. This indicates that among various biases, data bias stands out as particularly prevalent. Effectively addressing data bias is important while developing a fair AI/ML system.

\subsubsection{Can a fair AI/ML system ever be developed?}
As per the participants, a challenge they encountered in developing a fair AI/ML system revolved around \emph{obtaining the required datasets} for training the model. Several factors like `lack of cost', `lack of tools/techniques', `lack of representative datasets', and `lack of control over data collection' were reported that led to this challenge as discussed in Section \ref{sec:Resource-related challenge}. While factors like `lack of cost', `lack of tools/techniques', and `lack of control over data collection' could be addressable to improve the development of a fair AI/ML system, the majority of the participants reported the `lack of representative datasets' in the real world as one of the factors leading to the challenge of obtaining required datasets. For example, participants [P1] and [P15] said, 
\begin{quote}
    \small
\faCommenting \hspace{0.05cm} \emph{``In the real world, normally, we don't get perfect data to train the model.} - [P1]\\
\faCommenting \hspace{0.05cm} \emph{``So for example, in our human body projects, a lot of things are on bell curves in terms of like weight and height. And so it's often very difficult to get those data at the edges of the bell curve, you don't usually have a lot of very, very, like obese people."} - [P15]
\end{quote}

This particular factor, `the lack of representative datasets', appears to be more persistent because we cannot change real-world data, and it may pose a greater challenge that is not easily overcome. It was intriguing to learn that some participants believe developing a fair AI/ML system is not possible, asserting that while bias can be minimised, it cannot be entirely eradicated from the systems. For example, participants [P9] and [P11] said:
\begin{quote}
    \small
\faCommenting \hspace{0.05cm} \emph{``The real world is not perfect so you don't have all the datasets you need. So you won't be able to remove some sort of biases from them. You will have some sort of biases, the only thing that we can do is reduce it to a certain acceptable level. We won't be getting a perfect fair model. It's not there. Model reflects the data. Real data is not perfect. So you cannot expect a perfect model with it."} - [P9] \\
\faCommenting \hspace{0.05cm} \emph{``"In my view, because we are talking about bias here, like, there is no model I mean, even in AI, as far as I know, there is no perfect machine or model when prediction is involved. So in my opinion, I guess like a fair model should be something that decreases the bias, as you know, as I mean, it should decrease it. It should decrease it to be very less. But then I don't think there can be any model, which is not biased."} - [P11]
\end{quote}

Because only a small number of participants in our study talked about this subject, there is room for further investigation into why AI practitioners believe developing a perfectly fair (bias-free) AI/ML system is not feasible. 

\subsubsection{Organisational Impact vs. User Well-being}
We found that participants in our study believed that the repercussions of developing an unfair AI/ML system impact organisations more significantly than the users who interact with it. When asked about the consequences of developing an unfair AI/ML system, the majority highlighted potential financial losses and damage to the organisation's reputation. Participants expressed opinions such as \small\faCommenting \hspace{0.05cm} \emph{``it could ultimately tarnish the image of these organisations (Twitter)"} and \faCommenting \hspace{0.05cm} \emph{``incur a substantial loss for the company"}. \normalsize Only a small number of participants mentioned that users may experience emotional distress and discrimination when unfair AI/ML systems are developed. Participants quoted, \small \faCommenting \hspace{0.05cm} \emph{``the sentiments of the people can be hurt"}. \normalsize The limited acknowledgment of potential user experiences, such as emotional distress and discrimination, suggests a potential gap in awareness or consideration of the individual implications of unfair AI/ML systems. It may highlight a tendency to prioritise the broader consequences for organisations over the direct effects on the individuals interacting with these systems. 

\subsection{Implications}
This section outlines implications for researchers and AI practitioners involved in AI/ML system development, derived from our study findings. Additionally, we offer recommendations for AI practitioners and AI companies to assist in the development of fair AI/ML systems. 

\subsubsection{Implications for Research and Future Work}
We developed a framework to show the relationship between AI practitioners' \emph
{understanding} of `fair AI/ML' and the associated \emph{challenges} in developing a fair AI/ML system, the \emph{consequences} of developing an unfair AI/ML system perceived by them, and the \emph{strategies} employed to ensure fairness in AI/ML systems (Figure \ref{fig:Plot}) based on empirical findings. This framework can be used to identify patterns, and potential areas for intervention, ultimately contributing to a more nuanced understanding of how to enhance fairness in AI/ML systems. The insights drawn from this framework can inform future studies, shaping the direction of research in the field. Researchers can use our findings for future research in the following areas:

\faHandORight \hspace{0.05 cm}\textbf{Investigating factors and solutions for the challenge of obtaining required datasets:} Our findings reveal that a common challenge faced by AI practitioners who shared their understanding of `fair AI/ML' either in terms of the absence of \emph{bias} or in terms of the presence of desirable \emph{attributes} in AI/ML systems such as transparency, interpretability, accuracy etc. was obtaining necessary datasets during AI/ML system development, as discussed in Section \ref{def vs challenge}. Future work can focus on addressing more factors leading to this challenge and investigating approaches to mitigate them, which can contribute to the development of a fair AI/ML system.

\faHandORight \hspace{0.05 cm} \textbf{Mapping countries'/companies' definitions of `AI fairness' with practitioners' understanding:} Our findings reveal variations in AI practitioners' understanding of `fair AI/ML' compared to definitions set by different countries and tech companies on `AI fairness' (section \ref{sec:Def_Discussion}). Future research could explore the alignment between these perspectives through a mapping exercise.

\faHandORight \hspace{0.05 cm}\textbf{Delving deeper into strategies for ensuring fairness in AI/ML systems:} Our findings show that AI practitioners commonly use mitigating bias (bias-related strategy) to ensure fairness in AI/ML systems, regardless of how they describe `fair AI/ML', as discussed in Section \ref{def vs strategy}. Similarly, the participants discussed the strategy of detecting inaccuracy to ensure fairness in AI/ML systems; however, there is no mention of strategies to address accuracy-related issues for ensuring fairness. Future research can delve into why mitigating bias is the predominant strategy and explore if practitioners employ strategies to address accuracy-related issues in the system to ensure fairness. This may help to inform the development of comprehensive strategies to address both bias and accuracy concerns, leading to more robust and fair AI/ML systems. 

\faHandORight \hspace{0.05 cm}\textbf{Exploring links between various aspects:} In our study, we explored the link between what AI practitioners \emph{understand} by `fair AI/ML' and the \emph{challenges} they face in development, the \emph{consequences} of developing an unfair AI/ML systems perceived by them, and the \emph{strategies} they employed to ensure fairness of an AI/ML system. In the future, researchers can delve into exploring connections between other aspects, such as the challenges encountered in developing a fair AI/ML system and the consequences of developing an unfair AI/ML system perceived by AI practitioners. This may help to uncover deeper insights and connections within the complex landscape of developing a fair AI/ML system, guiding researchers in refining methodologies, devising more effective strategies, and advancing fair and ethical practices in AI/ML system development. 

\subsubsection{Implications for Practice and Recommendations}
Our study focuses on investigating AI practitioners' experiences and perceptions about various aspects related to the development of a fair AI/ML system. We conducted semi-structured interviews with 22 AI practitioners, exploring their \textbf{understanding} of `fair AI/ML', the \textbf{challenges} encountered in its development, \textbf{consequences} of developing an unfair AI/ML system, and the \textbf{strategies} employed to ensure the fairness of an AI/ML system. Our findings provide AI practitioners with valuable insights, into how people in the same field understand a `fair AI/ML', the challenges they encounter, the consequences of developing an unfair AI/ML system, and the strategies they employ to ensure fairness in an AI/ML system. This comprehensive understanding, derived from real-world experiences, can inform practitioners' approaches, enhance decision-making, and contribute to the use of more effective strategies for developing a fair AI/ML system. It may provide a practical and grounded perspective that can guide practitioners in navigating the complexities of fairness in their AI/ML development processes. 

Drawing from our study's findings, we present some recommendations for AI practitioners and AI companies to support the development of a fair AI/ML system, as detailed below.

\faLightbulbO \hspace{0.05 cm} \textbf{Recommendation 1: Striking a balance between the fairness of a system and its working version:} Several participants in our study highlighted the challenge of developing their envisioned ideal system, attributing it to factors like a shortage of time. Consequently, they prioritise creating a functional system over ensuring its fairness, as discussed in Section \ref{sec:process-related challenge}. AI managers can help AI practitioners by fostering a culture that values and prioritises fairness in AI/ML system development. They can allocate resources, both in terms of time and support, to enable practitioners to strike a balance between developing a working system and ensuring its fairness. 

\faLightbulbO \hspace{0.05 cm} \textbf{Recommendation 2: Providing AI practitioners with necessary tools/techniques:} Many participants in our study emphasised the challenges of developing a fair AI/ML system, citing a lack of tools, or techniques as discussed in Sections \ref{sec:process-related challenge} and \ref{sec:Resource-related challenge}. They specifically pointed out challenges in detecting and addressing data bias and obtaining necessary datasets due to the absence of adequate tools. AI companies can provide substantial support by investing in the development and provision of specialised tools, and/or techniques aimed at addressing the challenges highlighted by participants \citep{holstein2019improving}.

\faLightbulbO \hspace{0.05 cm} \textbf{Recommendation 3: Focusing on enhancing own knowledge and awareness of different concepts:} The majority of participants in our study acknowledged facing challenges in grasping the concepts of `bias' and `fairness' as discussed in Section \ref{sec: team-related challenge}. They attributed this challenge to a lack of awareness and knowledge about these concepts, as well as a deficit in understanding the domain they work in. AI practitioners can take proactive steps such as seeking additional training or education on the concepts of `bias' and `fairness.' Engaging in domain-specific learning to enhance their understanding of the context they work in might also be beneficial. Staying informed about the latest developments and best practices in AI fairness can contribute to a more comprehensive understanding of these concepts.

\faLightbulbO \hspace{0.05 cm} \textbf{Recommendation 4: Prioritising users in AI/ML system development:}
In discussions about the consequences of developing unfair AI/ML systems perceived by the participants, most participants focused on the negative impacts on organisations, including financial losses and reputational repercussions (Section \ref{sec:Discussion}). Interestingly, only a small number recognised the potential emotional distress and discrimination experienced by end users as a consequence of such systems. AI practitioners can make a conscious effort to shift the focus from solely considering organisational consequences to understanding the direct impact on users. This might allow them to identify and address potential biases and discriminatory outcomes, contributing to the development of fair AI/ML systems that treat users equitably. Involving users across various phases of AI/ML system development to gather feedback might help in ensuring user-centric AI/ML system development. A recent study highlights the need for increased user engagement throughout algorithmic development to enhance fairness in AI algorithms \citep{dankloff2024analysing}.

\faLightbulbO \hspace{0.05 cm} \textbf{Recommendation 5: Updating and adapting AI ethics policies in organisations:}
Participants in our study identified challenges in adhering to policies and regulations within their organisations, citing outdated AI ethics policies and a lack of adaptation as the factors leading to those challenges, as discussed in Section \ref{sec:process-related challenge}. To address this, AI companies can prioritise updating and adapting their AI ethics policies, ensuring strict adherence by practitioners. This proactive approach can help ensure that AI practitioners are equipped with the latest guidelines to navigate complex ethical challenges, promoting responsible AI development.

\section{Limitations and Threats to Validity} \label{sec:Limitations}
While we advertised our study on platforms such as LinkedIn and Twitter to attract participants globally, our data collection lacks an even distribution of participants worldwide. The majority of study participants are based in Australia. The findings of our study hold the most relevance for participants' organisations and their respective countries, potentially extending to similar contexts. However, generalising these findings to the entire global software engineering community is deemed impractical in practice \citep{masood2020agile}. The limitation of this study is that all interview participants held purely technical roles, such as AI/ML developers, engineers, experts, and data scientists involved in AI/ML system design and development. The study did not include a broader range of profiles like data science managers, business experts, ethics/compliance officers, risk managers, heads of innovation, and heads of operations. Including these profiles could have provided a wider range of perspectives on AI/ML fairness. Future studies should incorporate these roles to better understand their perspectives on AI/ML fairness.

Likewise, our main interview guide was developed after conducting two pilot interviews. The interview recordings underwent automatic transcription, and any errors introduced during this process were manually rectified by listening to each audio recording during the coding phase. In the interviews, there could be a possibility of misalignment between our intended questions and participants' understanding, leading to potential misinterpretations or misunderstandings. To address this, we employed follow-up questions to ensure clarity on the participants' statements.

All four authors were involved in designing the interview guide, with the initial coding primarily handled by the first author. However, all authors actively participated in refining and finalising the codes, concepts, and categories through collaborative discussions. We have also included various interview quotes as examples, aiming to minimise any potential reporting biases in the study. 

In addition, there could be a potential risk to the research’s internal validity when using the payment for the second round of data collection. As a way of mitigating this risk, we initially provided the candidates with an anonymous pre-interview questionnaire asking them about their years of experience in AI/ML system development. Using this information, we selected participants for interviews, and approval for payment was granted only after confirming alignment with our predetermined participation criteria. The process was carried out with ethics approval. Candidates with no experience in AI/ML system development were not selected for the interview.

\section{Conclusion} \label{sec:Conclusion}
This study aimed to investigate AI practitioners' perspectives and experiences in developing a fair AI/ML system, recognising their pivotal role in development and deployment. The study contributes to gaining insights into the industry's standpoint on the \emph{understanding} of a `fair AI/ML', the \emph{challenges} involved in its development, the \emph{consequences} of developing an unfair AI/ML system perceived by them, and the \emph{strategies} they employed to ensure fairness of an AI/ML system.

We conducted semi-structured interviews with 22 AI practitioners to fulfill the objective of our study and analysed the qualitative data using the \emph{STGT for data analysis} \citep{hoda2021socio}. The analysis revealed two categories of AI practitioners' \textbf{understanding} of `fair AI/ML' including, (i) \emph{In terms of the absence of bias} and (ii) \emph{In terms of the presence of desirable attributes} in AI/ML systems. We also categorised the \textbf{challenges} of the participants in developing a fair AI/ML system into three sections including, (i) \emph{Process-related challenges}, (ii) \emph{Resource-related challenges}, and (iii) \emph{Team-related challenges}. Similarly, our analysis showed three categories of negative \textbf{consequences} perceived by participants in developing an unfair AI/ML system: (i) \emph{Impact on organisations}, (ii) \emph{Impact on users}, and (iii) \emph{Impact on practitioners}. We also classified the \textbf{strategies} employed by participants to ensure the fairness of an AI/ML system into two categories: (i) \emph{Bias-related strategies} and (ii) \emph{Performance-related strategies}. Based on the findings, we also developed a framework to show the relationship between AI practitioners' \emph{understanding} of `fair AI/ML' and three other aspects including, (i) their \emph{challenges} in developing a fair AI/ML system, (ii) the \emph{consequences} of developing an unfair AI/ML system perceived by them and (iii) their \emph{strategies} to ensure the fairness of an AI/ML system.

Our findings offer valuable insights into the industry's perspective and experiences in developing a fair AI/ML system, aiding the AI research community in better understanding how AI practitioners perceive and experience this process. We also identified areas that need further investigation within the AI research community, enabling researchers to make more informed decisions about the direction of their studies. This might ensure that their efforts address the critical areas identified by the study for further exploration. We also offered recommendations to AI practitioners and AI companies, aiming to assist in enhancing the development of a fair AI/ML system.

\begin{acknowledgements}
Aastha Pant is supported by the Faculty of IT Ph.D. scholarship from Monash University. C. Tantithamthavorn is partially supported by the Australian Research Council’s Discovery Early Career Researcher Award (DECRA) funding scheme (DE200100941).
We would like to thank all the interviewees for their participation in our study. 

\end{acknowledgements}

\section{Appendices}
\appendix

\scriptsize
\section{Appendix A: Interview Protocol} \label{Appendix A}
\textbf{Section A: Demographic Information (via \textit{Qualtrics})}
\scriptsize
  1.  Your full name:
  
   2.  Please enter your email address so the researcher can contact you to schedule a time for an interview: 
   
    3. What is your current job title?
    \begin{itemize}
    \scriptsize
        \item AI Engineer
        \item AI/ML/Data Scientist
        \item AI/ML Expert
        \item AI/ML Practitioner
        \item AI/ML Developer
        \item Other:
            \end{itemize}

            4. How many years of experience do you have in the area of AI/ML system development?
            \begin{itemize}
                \scriptsize
                \item No Experience
                \item Less than 1 year
                \item Between 1 to 2 years
                \item Between 2 to 5 years
                \item More than 5 years
            \end{itemize}

       5. How old are you?
        \begin{itemize}
            \scriptsize
            \item Below 20
            \item 20-25
            \item 26-30
            \item 31-35
            \item 36-40
            \item 41-45
            \item 46-50
            \item 50+
        \end{itemize}

        6.  How would you describe your gender?
        \begin{itemize}
            \scriptsize
            \item Woman
            \item Man
            \item Non-binary/ gender diverse
            \item My gender identity isn't listed. I identify as:
            \item Prefer not to say
        \end{itemize}
        
7. What is your country of residence? 

8. What is the highest degree or level of education you have completed?
\begin{itemize}
    \scriptsize
    \item High School
\item Bachelor degree
\item Master degree
\item Ph.D. or Higher
\item Prefer not to answer
\item Other:
\end{itemize}

9. What activities are you involved in? Select \textbf{all} that apply.
\begin{itemize}
    \scriptsize
    \item Model requirements
\item Data collection
\item Data cleaning
\item Data labeling
\item Feature engineering
\item Model training
\item Model evaluation
\item Model deployment
\item Model monitoring
\item Other:
\end{itemize}

\textbf{Section B: Practitioners' Perception and Experiences on AI/ML Fairness (via semi-structured interviews)}\\

\textbf{Section B.1- Questions on `AI/ML Bias'}
\begin{enumerate}
    \scriptsize
\item  Can you briefly tell me about your professional background and current role?
\item Are you aware of the term `AI/ML bias’?
\begin{enumerate}
    \scriptsize
       \item (If yes), what do you understand by the term ‘AI/ML bias’?
\end{enumerate}
\item Based on your professional experience, can you tell me if something like ‘AI/ML bias’ exists in practice?

\begin{enumerate}
    \scriptsize
    \item (If yes), why do you say so? In your professional experience, have you come across any cases related to AI/ML bias?)
    \item (If yes), can you give an example?
    \item What kind of AI/ML system were you developing?
\item How did you find out that the system was biased?
\item What kind of bias crept into the system?
\item What caused the bias?
\item Once you found that the system was biased, did you deploy that system? (Yes/No)
\begin{enumerate}
    \scriptsize
   \item (If yes), any strategies/methods were used to mitigate those biases before
deploying it?
        \item(If yes), what strategies/methods did you use?
         \item (If not), why weren't any strategies/methods used?
\end{enumerate}
\item (If no), do you think the term ‘AI/ML bias’ is theoretical and does not exist in practice?
\begin{enumerate}
    \scriptsize
    \item You do not have to deal with/haven't dealt with any AI/ML biases? Can you tell me based on your experience?
\item Why don't you have to deal with AI/ML bias?
\end{enumerate}
\end{enumerate}
\item Based on your professional experience, what can help you in addressing/ preventing/ mitigating biases in the AI/ML system you develop?
\begin{enumerate}
    \scriptsize
    \item In what way can it help you?
\end{enumerate}
\end{enumerate}

\textbf{Section B.2- Questions on `Fair AI/ML'}
\begin{enumerate}
       \scriptsize
   \item Are you aware of the term `fair AI/ML'?

\begin{enumerate}
    \scriptsize
    \item (If yes), what would you consider as fair? Can you give an example?
\end{enumerate}
    \item Based on your professional experience, do you think it is important to create a fair AI/ML system?
    \begin{enumerate}
        \scriptsize
        \item (If yes), why is it important for the AI/ML system to be fair? 
        \item (If not), why is it not important to create fair AI/ML systems?
    \end{enumerate}

\item  Based on your professional experience, how do you know the AI/ML system that you developed is fair? Do you use any strategies to ensure its fairness?
    \begin{enumerate}
        \scriptsize
    \item (If yes), what strategies/ tools/ techniques do you use?
        \item (If not), is it not mandatory to use tools/strategies/techniques?
        \item Why is it not mandatory?
    \end{enumerate}

\item Based on your professional experience, do you face any challenges in developing a fair AI/ML system?
\begin{enumerate}
    \scriptsize
    \item (If yes), what challenges do you face?
    \item What do you think are the factors leading to those challenges? 
\end{enumerate}

\item Based on your professional experience, what does it take to develop AI/ML systems that are fair?
\begin{enumerate}
    \scriptsize
    \item  Why?
\end{enumerate}
\item If an AI/ML system is unfair, who does it impact, according to you?
\begin{enumerate}
    \scriptsize
    \item (If yes), how does it impact them? Can you give an example?
    \item (If not), why not? Can you give an example?
\end{enumerate}
\end{enumerate}

\appendix
\section*{Data Availability Statement}
The data are protected and are not available due to data privacy laws.

 \section*{Conflict of interest}
Conflicts of interest include Klaas-Jan Stol, Paul Ralph, Brian Fitzgerald, Burak Turhan, Patanamon Thongtanunam.

\bibliographystyle{spbasic} 
\bibliography{bibfile}

\begin{thebibliography}{62}
\providecommand{\natexlab}[1]{#1}
\providecommand{\url}[1]{{#1}}
\providecommand{\urlprefix}{URL }
\expandafter\ifx\csname urlstyle\endcsname\relax
  \providecommand{\doi}[1]{DOI~\discretionary{}{}{}#1}\else
  \providecommand{\doi}{DOI~\discretionary{}{}{}\begingroup \urlstyle{rm}\Url}\fi
\providecommand{\eprint}[2][]{\url{#2}}

\bibitem[{Angwin et~al.(2016)Angwin, Larson, Mattu, and Kirchner}]{Machine}
Angwin J, Larson J, Mattu S, Kirchner L (2016) Machine bias. \urlprefix\url{https://www.propublica.org/article/machine-bias-risk-assessments-in-criminal-sentencing}, accessed 17 January 2024

\bibitem[{Augustin et~al.(2002)Augustin, Bressler, and Smith}]{augustin2002accelerating}
Augustin L, Bressler D, Smith G (2002) Accelerating software development through collaboration. In: Proceedings of the 24th International Conference on Software Engineering, pp 559--563, \doi{https://doi.org/10.1145/581339.581409}

\bibitem[{Australia(2019)}]{Australisethics}
Australia G (2019) Australia’s \uppercase {AI} ethics principles. \urlprefix\url{https://www.industry.gov.au/publications/australias-artificial-intelligence-ethics-framework/australias-ai-ethics-principles}, accessed 10 January 2024

\bibitem[{Bacelar(2021)}]{bacelar2021monitoring}
Bacelar M (2021) Monitoring bias and fairness in machine learning models: A review. ScienceOpen Preprints \doi{10.14293/S2199-1006.1.SOR-.PP59WRH.v1}

\bibitem[{Balayn et~al.(2023)Balayn, Yurrita, Yang, and Gadiraju}]{balayn2023fairness}
Balayn A, Yurrita M, Yang J, Gadiraju U (2023) “\uppercase{F}airness toolkits, \uppercase{A} checkbox culture?” \uppercase{O}n the factors that fragment developer practices in handling algorithmic harms. In: Proceedings of the 2023 AAAI/ACM Conference on AI, Ethics, and Society, pp 482--495, \doi{https://doi.org/10.1145/3600211.3604674}

\bibitem[{Baltes and Ralph(2022)}]{baltes2022sampling}
Baltes S, Ralph P (2022) Sampling in software engineering research: A critical review and guidelines. Empirical Software Engineering 27(4):94, \doi{https://doi.org/10.1007/s10664-021-10072-8}

\bibitem[{Binns(2018)}]{binns2018fairness}
Binns R (2018) Fairness in machine learning: Lessons from political philosophy. In: Conference on Fairness, Accountability and Transparency, PMLR, pp 149--159

\bibitem[{Caliskan et~al.(2017)Caliskan, Bryson, and Narayanan}]{caliskan2017semantics}
Caliskan A, Bryson JJ, Narayanan A (2017) Semantics derived automatically from language corpora contain human-like biases. Science 356(6334):183--186, \doi{10.1126/science.aal42}

\bibitem[{Caton and Haas(2020)}]{caton2020fairness}
Caton S, Haas C (2020) Fairness in machine learning: A survey. ACM Computing Surveys 56(166):1--38, \doi{https://doi.org/10.1145/3616865}

\bibitem[{Chen et~al.(2023)Chen, Wu, and Wang}]{chen2023ai}
Chen P, Wu L, Wang L (2023) \uppercase{AI} fairness in data management and analytics: A review on challenges, methodologies and applications. Applied Sciences 13(18):10258, \doi{https://doi.org/10.3390/app131810258}

\bibitem[{Chouldechova and Roth(2018)}]{chouldechova2018frontiers}
Chouldechova A, Roth A (2018) The frontiers of fairness in machine learning. arXiv preprint arXiv:181008810

\bibitem[{D'Amour et~al.(2020)D'Amour, Srinivasan, Atwood, Baljekar, Sculley, and Halpern}]{d2020fairness}
D'Amour A, Srinivasan H, Atwood J, Baljekar P, Sculley D, Halpern Y (2020) Fairness is not static: Deeper understanding of long term fairness via simulation studies. In: Proceedings of the 2020 Conference on Fairness, Accountability, and Transparency, pp 525--534, \doi{https://doi.org/10.1145/3351095.3372878}

\bibitem[{Dankloff et~al.(2024)Dankloff, Skoric, Sileno, Ghebreab, Ossenbruggen, and Beauxis-Aussalet}]{dankloff2024analysing}
Dankloff M, Skoric V, Sileno G, Ghebreab S, Ossenbruggen Jv, Beauxis-Aussalet E (2024) Analysing and organising human communications for \uppercase{AI} fairness assessment: Use cases from the dutch public sector. AI \& Society pp 1--21, \doi{https://doi.org/10.1007/s00146-024-01974-4}

\bibitem[{Deng et~al.(2022)Deng, Nagireddy, Lee, Singh, Wu, Holstein, and Zhu}]{deng2022exploring}
Deng WH, Nagireddy M, Lee MSA, Singh J, Wu ZS, Holstein K, Zhu H (2022) Exploring how machine learning practitioners (try to) use fairness toolkits. In: Proceedings of the 2022 ACM Conference on Fairness, Accountability, and Transparency, pp 473--484, \doi{https://doi.org/10.1145/3531146.3533113}

\bibitem[{DrivenData(2024)}]{Deon}
DrivenData (2024) Deon. \urlprefix\url{https://deon.drivendata.org/}, accessed 17 January 2024

\bibitem[{Fenu et~al.(2022)Fenu, Galici, and Marras}]{fenu2022experts}
Fenu G, Galici R, Marras M (2022) Experts’ view on challenges and needs for fairness in artificial intelligence for education. In: International Conference on Artificial Intelligence in Education, Springer, pp 243--255, \doi{https://doi.org/10.1007/978-3-031-11644-5_20}

\bibitem[{Finkelstein et~al.(2008)Finkelstein, Harman, Mansouri, Ren, and Zhang}]{finkelstein2008fairness}
Finkelstein A, Harman M, Mansouri SA, Ren J, Zhang Y (2008) “\uppercase{F}airness analysis” in requirements assignments. In: 16th IEEE International Requirements Engineering Conference, IEEE, pp 115--124, \doi{10.1109/RE.2008.61}

\bibitem[{Friedler et~al.(2019)Friedler, Scheidegger, Venkatasubramanian, Choudhary, Hamilton, and Roth}]{friedler2019comparative}
Friedler SA, Scheidegger C, Venkatasubramanian S, Choudhary S, Hamilton EP, Roth D (2019) A comparative study of fairness-enhancing interventions in machine learning. In: Proceedings of the Conference on Fairness, Accountability, and Transparency, pp 329--338, \doi{https://doi.org/10.1145/3287560.3287589}

\bibitem[{Google(2022)}]{Google}
Google (2022) Responsible \uppercase{AI} practices. \urlprefix\url{https://ai.google/responsibility/responsible-ai-practices/}, accessed 10 January 2024

\bibitem[{Group(2019)}]{EUethics}
Group HLE (2019) Ethics guidelines for trustworthy \uppercase{AI}. \urlprefix\url{https://digital-strategy.ec.europa.eu/en/library/ethics-guidelines-trustworthy-ai}, accessed 10 January 2024

\bibitem[{Habibullah et~al.(2023)Habibullah, Gay, and Horkoff}]{habibullah2023non}
Habibullah KM, Gay G, Horkoff J (2023) Non-functional requirements for machine learning: Understanding current use and challenges among practitioners. Requirements Engineering 28(2):283--316, \doi{https://doi.org/10.1007/s00766-022-00395-3}

\bibitem[{Harrison et~al.(2020)Harrison, Hanson, Jacinto, Ramirez, and Ur}]{harrison2020empirical}
Harrison G, Hanson J, Jacinto C, Ramirez J, Ur B (2020) An empirical study on the perceived fairness of realistic, imperfect machine learning models. In: Proceedings of the 2020 Conference on Fairness, Accountability, and Transparency, pp 392--402, \doi{https://doi.org/10.1145/3351095.3372831}

\bibitem[{Hoda(2021)}]{hoda2021socio}
Hoda R (2021) Socio-technical grounded theory for software engineering. IEEE Transactions on Software Engineering 48(10):3808--3832, \doi{10.1109/TSE.2021.3106280}

\bibitem[{Holstein et~al.(2019)Holstein, Wortman~Vaughan, Daum{\'e}~III, Dudik, and Wallach}]{holstein2019improving}
Holstein K, Wortman~Vaughan J, Daum{\'e}~III H, Dudik M, Wallach H (2019) Improving fairness in machine learning systems: What do industry practitioners need? In: Proceedings of the 2019 CHI Conference on Human Factors in Computing Systems, pp 1--16, \doi{https://doi.org/10.1145/3290605.3300830}

\bibitem[{Hopkins and Booth(2021)}]{hopkins2021machine}
Hopkins A, Booth S (2021) Machine learning practices outside big tech: How resource constraints challenge responsible development. In: Proceedings of the 2021 AAAI/ACM Conference on AI, Ethics, and Society, ACM New York, United States, pp 134--145, \doi{https://doi.org/10.1145/3461702.3462527}

\bibitem[{Hua and Belfield(2020)}]{hua2020ai}
Hua SS, Belfield H (2020) \uppercase{AI} \& antitrust: Reconciling tensions between competition law and cooperative \uppercase{AI} development. Yale JL \& Tech 23:415

\bibitem[{Hutchinson and Mitchell(2019)}]{hutchinson201950}
Hutchinson B, Mitchell M (2019) 50 years of test (un) fairness: Lessons for machine learning. In: Proceedings of the Conference on Fairness, Accountability, and Transparency, ACM New York, USA, pp 49--58, \doi{https://doi.org/10.1145/3287560.3287600}

\bibitem[{IBM(2022)}]{IBM}
IBM (2022) Everyday ethics for \uppercase{AI}. \urlprefix\url{https://www.ibm.com/design/ai/ethics/everyday-ethics}, accessed 10 January 2024

\bibitem[{IBM(2024{\natexlab{a}})}]{AIFact}
IBM (2024{\natexlab{a}}) \uppercase{AI} factsheets. \urlprefix\url{https://dataplatform.cloud.ibm.com/docs/content/wsj/analyze-data/factsheets-model-inventory.html?context=cpdaas}, accessed 17 January 2024

\bibitem[{IBM(2024{\natexlab{b}})}]{AI360}
IBM (2024{\natexlab{b}}) \uppercase{AI} fairness 360. \urlprefix\url{https://www.ibm.com/opensource/open/projects/ai-fairness-360/}, accessed 17 January 2024

\bibitem[{Johnson and Brun(2022)}]{johnson2022fairkit}
Johnson B, Brun Y (2022) Fairkit-learn: A fairness evaluation and comparison toolkit. In: Proceedings of the ACM/IEEE 44th International Conference on Software Engineering: Companion Proceedings, pp 70--74, \doi{https://doi.org/10.1145/3510454.3516830}

\bibitem[{Madaio et~al.(2022)Madaio, Egede, Subramonyam, Wortman~Vaughan, and Wallach}]{madaio2022assessing}
Madaio M, Egede L, Subramonyam H, Wortman~Vaughan J, Wallach H (2022) Assessing the fairness of \uppercase{AI} systems: \uppercase{AI} practitioners' processes, challenges, and needs for support. Proceedings of the ACM on Human-Computer Interaction 6(CSCW1):1--26, \doi{https://doi.org/10.1145/3512899}

\bibitem[{Madaio et~al.(2020)Madaio, Stark, Wortman~Vaughan, and Wallach}]{madaio2020co}
Madaio MA, Stark L, Wortman~Vaughan J, Wallach H (2020) Co-designing checklists to understand organizational challenges and opportunities around fairness in \uppercase{AI}. In: Proceedings of the 2020 CHI Conference on Human Factors in Computing Systems, ACM New York, USA, pp 1--14, \doi{https://doi.org/10.1145/3313831.3376445}

\bibitem[{Majumder et~al.(2023)Majumder, Chakraborty, Bai, Stolee, and Menzies}]{majumder2023fair}
Majumder S, Chakraborty J, Bai GR, Stolee KT, Menzies T (2023) Fair enough: Searching for sufficient measures of fairness. ACM Transactions on Software Engineering and Methodology 32(6):1--22, \doi{https://doi.org/10.1145/3585006}

\bibitem[{Marcinkowski et~al.(2020)Marcinkowski, Kieslich, Starke, and L{\"u}nich}]{marcinkowski2020implications}
Marcinkowski F, Kieslich K, Starke C, L{\"u}nich M (2020) Implications of \uppercase{AI} (un-) fairness in higher education admissions: The effects of perceived \uppercase{AI} (un-) fairness on exit, voice and organizational reputation. In: Proceedings of the 2020 Conference on Fairness, Accountability, and Transparency, ACM New York, USA, pp 122--130, \doi{https://doi.org/10.1145/3351095.3372867}

\bibitem[{Martin(2018)}]{Amazon}
Martin N (2018) Are \uppercase{AI} hiring programs eliminating bias or making it worse? \urlprefix\url{https://www.forbes.com/sites/nicolemartin1/2018/12/13/are-ai-hiring-programs-eliminating-bias-or-making-it-worse/?sh=552bb0cc22b8}, accessed 17 January 2024

\bibitem[{Masood et~al.(2020)Masood, Hoda, and Blincoe}]{masood2020agile}
Masood Z, Hoda R, Blincoe K (2020) How agile teams make self-assignment work: A grounded theory study. Empirical Software Engineering 25:4962--5005, \doi{https://doi.org/10.1007/s10664-020-09876-x}

\bibitem[{Mehrabi et~al.(2021)Mehrabi, Morstatter, Saxena, Lerman, and Galstyan}]{mehrabi2021survey}
Mehrabi N, Morstatter F, Saxena N, Lerman K, Galstyan A (2021) A survey on bias and fairness in machine learning. ACM Computing Surveys (CSUR) 54(6):1--35, \doi{https://doi.org/10.1145/3457607}

\bibitem[{Microsoft(2024{\natexlab{a}})}]{Microsoft}
Microsoft (2024{\natexlab{a}}) Microsoft responsible \uppercase{AI} standard. \urlprefix\url{https://www.microsoft.com/en-us/ai/responsible-ai?activetab=pivot1\%3aprimaryr6}, accessed 10 January 2024

\bibitem[{Microsoft(2024{\natexlab{b}})}]{Microsoft_checklist}
Microsoft (2024{\natexlab{b}}) \uppercase{AI} fairness checklist. \urlprefix\url{https://www.microsoft.com/en-us/research/project/ai-fairness-checklist/}, accessed 17 January 2024

\bibitem[{Orr and Davis(2020)}]{orr2020attributions}
Orr W, Davis JL (2020) Attributions of ethical responsibility by artificial intelligence practitioners. Information, Communication \& Society 23(5):719--735, \doi{https://doi.org/10.1080/1369118X.2020.1713842}

\bibitem[{Pagano et~al.(2023)Pagano, Loureiro, Lisboa, Peixoto, Guimar{\~a}es, Cruz, Araujo, Santos, Cruz, Oliveira et~al.}]{pagano2023bias}
Pagano TP, Loureiro RB, Lisboa FV, Peixoto RM, Guimar{\~a}es GA, Cruz GO, Araujo MM, Santos LL, Cruz MA, Oliveira EL, et~al. (2023) Bias and unfairness in machine learning models: A systematic review on datasets, tools, fairness metrics, and identification and mitigation methods. Big Data and Cognitive Computing 7(1):15, \doi{https://doi.org/10.3390/bdcc7010015}

\bibitem[{Pant et~al.(2023)Pant, Hoda, Spiegler, Tantithamthavorn, and Turhan}]{pant2023ethics}
Pant A, Hoda R, Spiegler SV, Tantithamthavorn C, Turhan B (2023) Ethics in the age of \uppercase{AI}: An analysis of \uppercase{AI} practitioners’ awareness and challenges. ACM Transactions on Software Engineering and Methodology 33(80):1--35, \doi{https://doi.org/10.1145/3635715}

\bibitem[{Pant et~al.(2024)Pant, Hoda, Turhan, and Tantithamthavorn}]{pant2024aimlpractitionersthinkaiml}
Pant A, Hoda R, Turhan B, Tantithamthavorn C (2024) What do \uppercase{AI/ML} practitioners think about \uppercase{AI/ML} bias? \urlprefix\url{https://arxiv.org/abs/2407.08895}, \eprint{2407.08895}

\bibitem[{Pessach and Shmueli(2022)}]{pessach2022review}
Pessach D, Shmueli E (2022) A review on fairness in machine learning. ACM Computing Surveys (CSUR) 55(3):1--44, \doi{https://doi.org/10.1145/3494672}

\bibitem[{Prates et~al.(2020)Prates, Avelar, and Lamb}]{prates2020assessing}
Prates MO, Avelar PH, Lamb LC (2020) Assessing gender bias in machine translation: A case study with google translate. Neural Computing and Applications 32:6363--6381, \doi{https://doi.org/10.1007/s00521-019-04144-6}

\bibitem[{Richardson et~al.(2021)Richardson, Garcia-Gathright, Way, Thom, and Cramer}]{richardson2021towards}
Richardson B, Garcia-Gathright J, Way SF, Thom J, Cramer H (2021) Towards fairness in practice: A practitioner-oriented rubric for evaluating fair \uppercase{ML} toolkits. In: Proceedings of the 2021 CHI Conference on Human Factors in Computing Systems, ACM New York, USA, pp 1--13, \doi{https://doi.org/10.1145/3411764.3445604}

\bibitem[{Ryan et~al.(2023)Ryan, Nadal, and Doherty}]{ryan2023integrating}
Ryan S, Nadal C, Doherty G (2023) Integrating fairness in the software design process: An interview study with \uppercase{HCI} and \uppercase{ML} experts. IEEE Access 11:29296--29313, \doi{10.1109/ACCESS.2023.3260639}

\bibitem[{Seaman(1999)}]{seaman1999qualitative}
Seaman CB (1999) Qualitative methods in empirical studies of software engineering. IEEE Transactions on Software Engineering 25(4):557--572, \doi{10.1109/32.799955}

\bibitem[{Shin and Park(2019)}]{shin2019role}
Shin D, Park YJ (2019) Role of fairness, accountability, and transparency in algorithmic affordance. Computers in Human Behavior 98:277--284, \doi{https://doi.org/10.1016/j.chb.2019.04.019}

\bibitem[{Silberg and Manyika(2019)}]{silberg2019notes}
Silberg J, Manyika J (2019) Notes from the \uppercase{AI} frontier: Tackling bias in \uppercase{AI} (and in humans). McKinsey Global Institute 1(6):1--31

\bibitem[{Srivastava et~al.(2019)Srivastava, Heidari, and Krause}]{srivastava2019mathematical}
Srivastava M, Heidari H, Krause A (2019) Mathematical notions vs. human perception of fairness: A descriptive approach to fairness for machine learning. In: Proceedings of the 25th ACM SIGKDD International Conference on Knowledge Discovery \& Data Mining, ACM New York, USA, pp 2459--2468, \doi{https://doi.org/10.1145/3292500.3330664}

\bibitem[{Ueda et~al.(2024)Ueda, Kakinuma, Fujita, Kamagata, Fushimi, Ito, Matsui, Nozaki, Nakaura, Fujima et~al.}]{ueda2024fairness}
Ueda D, Kakinuma T, Fujita S, Kamagata K, Fushimi Y, Ito R, Matsui Y, Nozaki T, Nakaura T, Fujima N, et~al. (2024) Fairness of artificial intelligence in healthcare: Review and recommendations. Japanese Journal of Radiology 42(1):3--15, \doi{https://doi.org/10.1007/s11604-023-01474-3}

\bibitem[{Vasudevan and Kenthapadi(2020)}]{vasudevan2020lift}
Vasudevan S, Kenthapadi K (2020) Lift: A scalable framework for measuring fairness in \uppercase{ML} applications. In: Proceedings of the 29th ACM International Conference on Information \& Knowledge Management, ACM New York, USA, pp 2773--2780, \doi{https://doi.org/10.1145/3340531.3412705}

\bibitem[{Verma and Rubin(2018)}]{verma2018fairness}
Verma S, Rubin J (2018) Fairness definitions explained. In: Proceedings of the International Workshop on Software Fairness, ACM New York, USA, pp 1--7, \doi{https://doi.org/10.1145/3194770.3194776}

\bibitem[{Wan et~al.(2023)Wan, Zha, Liu, and Zou}]{wan2023processing}
Wan M, Zha D, Liu N, Zou N (2023) In-processing modeling techniques for machine learning fairness: A survey. ACM Transactions on Knowledge Discovery from Data 17(3):1--27, \doi{https://doi.org/10.1145/3551390}

\bibitem[{Wang et~al.(2023)Wang, Song, Ma, and Han}]{wang2023multidisciplinary}
Wang Y, Song Y, Ma Z, Han X (2023) Multidisciplinary considerations of fairness in medical \uppercase{AI}: A scoping review. International Journal of Medical Informatics 178:105175, \doi{https://doi.org/10.1016/j.ijmedinf.2023.105175}

\bibitem[{Weidener et~al.(2024)Weidener, Fischer et~al.}]{weidener2024role}
Weidener L, Fischer M, et~al. (2024) Role of ethics in developing \uppercase{AI}-based applications in medicine: Insights from expert interviews and discussion of implications. JMIR AI 3(1):e51204, \doi{10.2196/51204}

\bibitem[{Woodruff et~al.(2018)Woodruff, Fox, Rousso-Schindler, and Warshaw}]{woodruff2018qualitative}
Woodruff A, Fox SE, Rousso-Schindler S, Warshaw J (2018) A qualitative exploration of perceptions of algorithmic fairness. In: Proceedings of the 2018 CHI Conference on Human Factors in Computing Systems, ACM New York, USA, pp 1--14, \doi{https://doi.org/10.1145/3173574.3174230}

\bibitem[{Xavier(2024)}]{xavier2024biases}
Xavier B (2024) Biases within \uppercase{AI}: Challenging the illusion of neutrality. AI \& Society pp 1--2, \doi{https://doi.org/10.1007/s00146-024-01985-1}

\bibitem[{Xivuri and Twinomurinzi(2021)}]{xivuri2021systematic}
Xivuri K, Twinomurinzi H (2021) A systematic review of fairness in artificial intelligence algorithms. In: Responsible AI and Analytics for an Ethical and Inclusive Digitized Society, Springer, vol 12896, pp 271--284, \doi{https://doi.org/10.1007/978-3-030-85447-8_24}

\bibitem[{Zhang et~al.(2023)Zhang, Shu, and Yu}]{zhang2023fairness}
Zhang J, Shu Y, Yu H (2023) Fairness in design: A framework for facilitating ethical artificial intelligence designs. International Journal of Crowd Science 7(1):32--39, \doi{10.26599/IJCS.2022.9100033}

\end{thebibliography}

\end{document}